\documentclass[twocolumn,showpacs,preprintnumbers,amsmath,amssymb]{revtex4}
\usepackage{graphicx}
\usepackage{dcolumn}
\usepackage{bm}

\def \b{\beta}        
\def \e{\epsilon}

\def\be{\begin{equation}}    \def\ee{\end{equation}}

\def \vp{{\bf p}}  
\def \vq{{\bf q}}  
\def \vk{{\bf k}}  

\begin{document}
\title{Expansion of the Gibbs potential for quantum many-body systems: General formalism
with applications  to the  spin glass  and the weakly non-ideal
Bose gas
  }
\author{T. Plefka}
\email{timm@fkp.tu-darmstadt.de} \affiliation{Department of Solid
State Physics , TU Darmstadt, D 64289 Darmstadt, Germany}
\date{\today}
\begin{abstract}
For  general quantum systems the power expansion of the Gibbs
potential and consequently the power expansion of the self  energy
is derived in terms of the interaction strength. Employing a
generalization of the projector technique a compact representation
of the general terms of the expansion results. The general aspects
of the approach are discussed with special emphasis on the effects
characteristic for quantum systems. The expansion is systematic
and leads directly to contributions beyond mean-field  of all
thermodynamic quantities. These features are explicitly
demonstrated and illustrated for two non-trivial  systems, the
infinite range quantum spin glass and the weakly interacting Bose
gas. The Onsager terms of  both systems are calculated, which
represent the first beyond mean-field contributions. For the spin
glass new TAP-like equations are presented and discussed in the
paramagnetic region. The investigation of the Bose gas leads to a
beyond mean-field  thermodynamic description. At the Bose-Einstein
condensation temperature  complete agreement is found with the
results presented recently by alternative techniques.
\end{abstract}
 \maketitle
\section{Introduction}
The problem of understanding the static properties of systems
possessing large or infinite numbers of particles pervades all of
theoretical physics. Apart from very rare exceptions,
approximation must be employed to find the characteristic features
of such many-particle systems. It is the static mean-field
approximation which is usually used to find  first descriptions of
these many-body systems. Although this approximation leads in many
cases to  reasonable descriptions there are situation where beyond
mean-field approximations are needed.

More than two decades ago, the present author  developed a method
\cite{I} to derive such beyond mean-field approximations in a
natural way. The approach is based on the power expansion of the
Gibbs potential and was performed for the Sherrington-Kirkpatrick
spin glass (SK) model \cite{sk}. The investigation confirmed  the
Thouless-Anderson-Palmer equations (TAP) that had been obtained
previously \cite{tap}. Due to the presence of infinite range
interactions,  the power expansion truncates. It are just the
terms up to the second order which  contribute in the
thermodynamic limit.

In subsequent publications, the method of \cite{I} was
successfully  applied   to the infinite ranged classical vector
spin glass by Vulovic \cite{vul}, to various other  spin glasses
\cite{other,opp} and to dynamical problems in the field of spin
glasses by Biroli and Cugliandolo \cite{dyn,bicu}. Neural networks
are closely related to spin glass \cite{mpv,fh,nishi}. Therefore
it is natural that the Gibbs potential approach has been used for
problems in this and related fields \cite{neural}. In particular
the first, generally accepted TAP-like approach to  the Hopfield
model, worked out by Nakanishi and Takayama \cite{nata-hop}, is
based on the method of \cite{I}.

The study of non spin glass-like problems with the Gibbs potential
approach   started with the contributions of Georges and Yedidia
for the ferromagnetic Ising and spherical models \cite{gyis} and
for the Hubbard model \cite{gyhub}. In addition to these
investigations, further work for non spin glass-like problems
\cite{non}  exists  that uses the method developed in \cite{I}. In
general, infinite range interactions are not presumed and
therefore a truncated expansion implies an approximation and
higher order terms may become important. For the general Ising
model, the third and forth order terms have  been calculated first
in \cite{gyis} and later in \cite{nata-hop}. Moreover, for some
special Ising or spherical systems,   leading terms beyond the
forth order  have been presented in these papers.

Note that the majority of the systems to which the expansion of
the Gibbs potential has been applied are classical systems.
Exceptions  are the investigations  for the Hubbard model
\cite{gyhub}, the fermionic spin glass model \cite{opp}  and the
quantum version of the spherical p spin glass model \cite{bicu}.
Apart from the fact that these approaches yield interesting
results, none of them is completely  representative for a
generalization of the Gibbs potential expansion to quantum
systems. It is just the Lagrange parameter conjugate to the order
parameter and the chemical potential for which the Legendre
transformation is performed in \cite{gyhub} for the Hubbard
Hamiltonian. In general,  the transformation is performed for a
larger set of variables. Thus the  approach  is very special and
therefore non-generic for quantum systems. All operators - the
spin operators, the  number operators and the interaction
Hamiltonian - of the fermionic spin glass model \cite{opp}
commute. Again  such a situation is  not representative for a
quantum system. Finally the approach \cite{bicu} uses very special
quantum variables and it is not obvious how to generalize this
work to obtain explicit results for usual quantum spins.

Hence, it is the aim of this work to present  a compete quantum
version of the power expansion for the Gibbs potential. To work
out the characteristic effects for quantum systems the approach
should be as general as possible. Thus we start in Sec.\ref{sgen}
with an arbitrary Hamiltonian and work out all terms of the
expansion up to an arbitrary order. In Sec.\ref{secdis} the formal
results are discussed from a general point of view.

The remaining part of this paper is reserved for  two specific
applications. In Sec.\ref{secspin} the infinite ranged  quantum
spin $ s=1/2 $ glass is treated for  non isotropic interactions.
TAP-like equations result  which are discussed for isotropic
interactions in the paramagnetic regime and compared to
corresponding equations for classical vector spins. In
Sec.\ref{secBose} the expansion of the Gibbs potential is worked
out for the weakly non-ideal Bose gas up the first beyond
mean-field contribution. The resulting equations of state are
valid in the entire temperature regime. For the self energy a
complete agreement with previous and alternative approaches is
found at the Bose-Einstein condensation temperature. In
Sec.\ref{seccon} we present some concluding remarks and compare to
other approaches.

Some introductory  remarks  should be added on the physical
systems to which the general results are applied. The quantum spin
$ s=1/2 $ glass with infinite range interactions is the natural
generalization of the SK or of the classical vector spin glass
model to include quantum effects. Despite being formulated over
two decades ago \cite{bmquant}, an understanding of this  spin
glass model has proven elusive \cite{parra}.

Although  the  theory of the weakly interacting Bose gas has a
long history,  some old  problems - like the effect on the
critical temperature due to such interactions - have been revived
(compare \cite{haque} for an overview). In this context detailed
beyond mean-field investigations for the self energy have recently
been published \cite{baym}.  As these results were derived  with
both Green's function methods and within the frameworks of  Ursell
operators \cite{ursell}, this approach is an ideal reference
system to compare the Gibbs potential approach with other work.

\section{General Formalism \label{sgen}}
\subsection{Basic concepts of the Gibbs potential expansion\label{s2a}}
The many-particle system is described by the thermodynamic
Hamiltonian ($\beta= 1/T$ and $ k_B=1$)
\begin{equation}\label{1} {\cal K}_\alpha= -\beta{\cal H} = \sum_i \nu^i_\alpha
{\cal A}^i + \alpha{\cal K}'
\end{equation}
where the single particle contributions and the many-particle
interactions correspond to $ \sum_i \nu^i_\alpha {\cal A}^i  $ and
to $ {\cal K}'$, respectively. It is assumed that the
thermodynamic quantities  and,  in particular, all expectation
values can be calculated for the free Hamiltonian ${\cal
K}_{\alpha=0}$ and that the  problem is to find at least
approximations for the interacting system.

 This is a very common
question and typical realizations are: interacting systems of
identical particles, classical or quantum spin systems and
combinations of these systems. The present approach to these
problems is quite general and does not need any further
requirements or assumptions.

The parameters $\nu^i_\alpha=(\nu^i_\alpha)^*$ are real and the
operators $ {\cal A}^i=({\cal A}^i)^\dag$ are Hermitian
\cite{com1}.   The variable $\alpha$ represents an (in general
complex) expansion parameter and the $ \nu^i_\alpha$ exhibit a
$\alpha$-dependence which will  be  specified below. The
Hamiltonian of the original problem corresponds to the value
$\alpha=1$. This implies that the values $ \nu^i_{\alpha=1} $ are
given and fixed by the specific physical problem under
investigation. Clearly this also implies that  we have to set
$\alpha=1$ at the end of the calculation.

The index $i$ is a shorthand index for both different particles
and different operators acting in the same subspace of the
individual particles. For a Hamiltonian in second quantization the
$ {\cal A}^i $ represent products of creation and destruction
operators and the index $i$ may become a pair index.

The object is to calculate a thermodynamic potential that
determines the relevant thermal mean values
\begin{equation}\label{2}
\big\langle \ldots\big \rangle_\alpha= \mathrm{ Tr}\ldots{\cal
R}_\alpha
\end{equation}
where the density operator  ${\cal R}_\alpha$ and the partition
function $ Z_\alpha$ are given by
\begin{equation}\label{3}
{\cal R}_\alpha =\frac{e^{{\cal K}_\alpha }}{Z_\alpha} \quad
\textrm{and\:by} \quad Z_\alpha =\mathrm{ Tr}\, e^{{\cal
K}_\alpha}\;,
\end{equation}
 respectively. The usual choice for  such a potential is the
free energy which is proportional to $ \ln Z_\alpha $. This work,
however, is focusing on the Gibbs potential $G_\alpha$ which is
related to the free energy by a Legendre transformation and
defined as
\begin{equation} \label{4}
G_\alpha(A^i) = \ln \, Z_\alpha \, -\sum_i \nu^i_\alpha A^i \quad
\mathrm{with}\quad A^i=\langle {\cal A}^i  \rangle_\alpha \quad.
\end{equation}
Strictly, it is the quantity $ -\beta G_\alpha(A^i)$ which
represents the thermodynamic Gibbs potential. We use, however, the
term for $ G_\alpha(A^i)$, keeping in mind this difference. For
systems  with  variable number of particles ${\cal R}_\alpha $ and
$Z_\alpha$ represent the grand canonical  density operators and
the grand canonical partition function , respectively. This
implies the slide modification  ${\cal K}_\alpha= -\beta \left
({\cal H} -\mu {\cal N}\right)$ for these systems where $\mu$ and
${\cal N}$ represent the chemical potential and the number
operator, respectively.

 The total differential of $G_\alpha$ is given by
\begin{equation}\label{400}
 \textrm{d} G_\alpha = \langle {\cal K}'\rangle_\alpha \textrm{d} \alpha
-\sum_i \nu^i_\alpha \textrm{d}
 A^i\quad.
\end{equation}
Thus the natural variables are the expansion parameter $\alpha$
and  the variables $A^i$ which are conjugate to the Lagrange
parameters $\nu^i_\alpha $. Thus both the Gibbs potential and the
Lagrange parameters $\nu^i_\alpha $ are functions of  $\alpha$ and
$A^i$ and the present approach exclusively uses  these natural
variables as independent variables.

Next $ G_\alpha(A^i)$ is expanded around $\alpha=0$ keeping the
values $A^i$ fixed to their thermal values. Thus the relation
$A^i=\langle {\cal A}^i \rangle_\alpha =\langle {\cal A}^i
\rangle_0=\langle {\cal A}^i \rangle_1$ holds. These constrains
determine the functions $\nu^i_\alpha(A^j)$ and remove the
arbitrariness of the above.  The Taylor expansion  leads to
\begin{equation}\label{5}
 G_\alpha= S_0 + \sum_{n=1}^\infty\frac{\alpha^n G^{(n)}}{n!}
\quad {\rm with} \quad G^{(n)}=  \frac{\partial^n
G_\alpha}{\partial
 \alpha^n}\Big{ |}_{\alpha=0}\:.
\end{equation}
The zeroth-order term $ S_0(A^i)$ of the expansion (\ref{5}) is
the entropy of the noninteracting system. This is easily checked
using Eq.(\ref{3}), the general definition  $ S_0 =- \langle\ln
{\cal R}_0 \rangle_0$ and
\begin{equation}\label{5000}
 {\cal K}_{0}= \sum \nu^i_0{\cal A}^{i}.
\end{equation}

Considering next the first-order contribution $G^{(1)}$
Eq.(\ref{400}) yields immediately
\begin{equation}\label{5001}
G^{(1)}= \langle {\cal K}'\rangle_0
\end{equation}
which represents nothing else as the usual mean-field energy. Thus
\textit{approximating the expansion (\ref{5})  by the first two
terms}  $\; G_{\alpha=1} \approx S_0 + \langle {\cal K}'\rangle_0$
\textit{corresponds to the standard mean-field theory.}

Differentiation of Eq.(\ref{5}) with respect to $ A^i$ and  using
Eq.(\ref{400}) yields  for $\alpha=0$
\begin{equation}\label{6000}
     \nu^i_0(A^i)   =-
\frac{\partial S_0}{\partial A^i  }
\end{equation}
and for $\alpha=1$
\begin{equation}\label{6}
     \nu^i_1   =\nu^i_0
 + \beta \Sigma^i \quad
\textrm{with}\: -\beta \Sigma^i =
\sum_{n=1}^\infty\frac{1}{n!}\frac{\partial G^{(n)}}{\partial
A^i}\:.
\end{equation}
According to Eq.(\ref{6000}) the values of the Lagrange parameters
$ \nu^i_0 $ are determined  by the derivative of the entropy of
the non interacting system that is a function of the $A_i $. This
relation is important. Indeed, Eq.(\ref{6000}) is   used  to
eliminate the dummy variables $ \nu^i_0 $  which enter in the
expectation values of the non interacting system via ${\cal
K}_{0}$ (compare Eq.(\ref{5000})).

Recall that the quantities $\nu^i_1$ represent the given parameter
values of the original Hamiltonian (\ref{1})  and note that both
terms $\nu^i_0 $ and $\Sigma^i$ are functions of the $A^i$.
Therefore the first Eq. of (\ref{6}) represents a \textit{thermic
 equations of states} from which the expectation values $A^i$ can be
determined for given values of the $\nu^i_1$ .

The second equation of (\ref{6}) represents the general
\textit{definition of the self- energy } $\Sigma^i$ or for
magnetic systems the general \textit{definition of the internal
field }. Note that this fact is well known in the theory of
Green's function where the Gibbs potential is often named
effective potential \cite{neor} . Thus the expansion of the Gibbs
potential implies a systematic expansion of the self-energy that
describes the effects of the interaction.

For specific examples these results are illustrated below and the
reader is referred to  Sec.\ref{secspin1} and to
Sec.\ref{secBose1} for magnetic systems and for a Bose gas,
respectively.

In this context it is remarked that a further differentiation of
Eq.(\ref{6}) with respect to $A_j$ results in an expansion of the
inverse susceptibility  matrix. This quantity is of high
importance as it governs the convergence of the expansion and
consequently the stability of the considered system - compare
\cite{I,II} for the SK model, Eq.(\ref{s25113}) for the quantum
spin glass and Eq.(\ref{b12}) for the Bose gas.

Rewriting  the definition of the coefficients $G^{(n)}$ as
\begin{equation}\label{5cum}
G^{(n)}=  \frac{\partial^n }{\partial \alpha^n}\ln\:\mathrm{Tr}
\exp{\Big \{\sum_i \nu^i_\alpha ({\cal A}^i-A_i) + \alpha{\cal
K}'}\Big \}\Big{ |}_{\alpha=0}
\end{equation}
it is obvious that the $G^{(n)}$ can be interpreted as a
generalization of the cumulant expectation values
\cite{kubo,fulde}. It is  just the additional $\alpha$ dependence
of the Lagrange parameters $\nu^i_\alpha $ that causes the
difference to  usual cumulants.

\subsection{ Onsager term}
In this subsection the explicit expression for  second-order term
$G^{(2)}=\partial_{\alpha\alpha} G_{\alpha\rightarrow 0} $  is
investigated. This term is called the Onsager term and represents
the lowest beyond mean-field contribution to the expansion.

According to Eq.(\ref{400}) $\partial_{\alpha} G_{\alpha}= \langle
{\cal K}'\rangle_\alpha $ holds which implies $
\partial_{\alpha\alpha} G_{\alpha} = {\rm Tr}\,{\cal
K}'\partial_{\alpha}{\cal R}_\alpha $. Focusing therefore  on the
calculation of $\partial_{\alpha}{\cal R}_\alpha $ the definitions
\begin{equation}\label{11}
  \mathbb{E}_\alpha \{{\cal
U} \} = \int_0^1 \,{\cal U}(\lambda)\,{\rm d} \lambda\ \quad ,
\quad {\cal U}(\lambda)= e^{\lambda{\cal K}_{\alpha}}{\cal U}
e^{-\lambda{\cal K}_{\alpha}}
\end{equation}
and
\begin{equation}\label{401}
({\cal U}|{\cal V})_\alpha=\langle {\cal
     U}^\dag\mathbb{E}_\alpha \{{\cal
     V} \}\rangle_\alpha
\end{equation}
are introduced. The definition (\ref{401}) represents the well
known Mori product \cite{mori} of the operators ${\cal U} $ and
${\cal V}$. This product is a scalar product in the Liouville
space, has many additional properties (compare Appendix
\ref{app0}) and is physically significant in particular for the
linear response theory (see \cite{fs} for a general reference).

The definition (\ref{11}) permits us to represent the differential
rule for exponential operators as
\begin{equation}\label{12}
{\rm d}\, e^{{\cal K}_{\alpha}}= \mathbb{E}_\alpha\{{\rm d\,}
{\cal K}_{\alpha}\}e^{{\cal K}_{\alpha}}\quad ; \quad {\rm d}
{\cal K}_\alpha={\cal K}'{\rm d}\alpha +\sum_i {\cal A}^{i}{\rm
d}\nu_\alpha^i .
\end{equation}
Due to the possibility of cyclic permutations within the trace
Eq.(\ref{12}) yields ${\rm d} Z_\alpha=Z_\alpha \langle {\rm d}
{\cal K}_{\alpha}\rangle_\alpha $ . This leads to the total
differential of the density operator
\begin{equation}\label{13}
 {\rm d} {\cal R}_\alpha=\mathbb{E}_\alpha\{\widetilde{{\cal K}}'\}{\cal R}_\alpha\,{\rm
d}\alpha +\sum_i \mathbb{E}_\alpha\{\widetilde{{\cal A}}^i\}{\cal
R}_\alpha\,{\rm d}\nu_\alpha^i
\end{equation}
where $\widetilde{{\cal U}}$ is defined by
\begin{equation}\label{14}
\widetilde{{\cal U}}={\cal U } -\langle{\cal
U}\rangle_\alpha\quad.
\end{equation}
 The constrains $ A^i= const. $ imply ${\rm Tr }\,{\cal A}^i{\rm
d} {\cal R}_\alpha=0 $. This leads to
\begin{equation}\label{15}
\partial_\alpha \nu_\alpha^i= -\sum_j\,
\Gamma^{ij}_\alpha \,( \widetilde{{\cal A}}^j |{\cal K}')_\alpha
\end{equation}
using the relation (\ref{aaa4}). The matrix $\Gamma^{ij}_\alpha$
is the inverse of the susceptibility matrix $\chi^{ij}_\alpha$
\begin{equation}\label{16}
\sum_k \, \Gamma^{ik}_\alpha \chi^{kj}_\alpha= \delta_{i j} \quad
\textrm{where} \quad \chi^{ij}_\alpha=( \widetilde{{\cal A}}^i
|\widetilde{{\cal A}}^j)_\alpha \quad.
\end{equation}
As the Mori product is a scalar product the matrix
$\chi^{ij}_\alpha$ is positive definite and thus the inverse
matrix $\Gamma^{ij}_\alpha$ exists.

For a compact notation it is convenient  to introduce the
projectors by $ \mathbb{P}_\alpha$ and $ \mathbb{Q}_\alpha$
\begin{equation}\label{17}
\mathbb{P}_\alpha  {\cal U}= (1|{\cal U})_\alpha + \sum_{ij}\,
\widetilde{{\cal A}}^i\,\Gamma_\alpha^{ij}\,( \widetilde{{\cal
A}}^j|{\cal U})_\alpha \quad;\quad \mathbb{Q}_\alpha
=\bm{1}-\mathbb{P}_\alpha \,.
\end{equation}
$\mathbb{P}_\alpha $ and $\mathbb{Q}_\alpha $ are super-operators
which  linearly map operators of the Hilbert-space onto other
operators of the Hilbert-space. With the above definitions it is
easy to show that the usual projector relations
$\mathbb{P}_\alpha^2=\mathbb{P}_\alpha$ ,
$\mathbb{Q}_\alpha^2=\mathbb{Q}_\alpha$ and $\mathbb{P}_\alpha
\mathbb{Q}_\alpha=0 $ are satisfied. The projector
$\mathbb{P}_\alpha$ projects onto the subspace that is spanned by
the elements ${\cal A}^i$ and by the unit operator $1$. These
bases elements are linearly independent but are in general not
orthogonal.

Let us introduce some definitions. In accordance with \cite{fs} we
use the term observation level for the set of operators $ {\cal
A}^i$   spanning  the subspace $\mathbb{P}_0$ together with the
unit operator. The set of all $  A^i$ and the set of all
$\nu^i_\alpha $ are called constrained  and conjugate variables,
respectively.

Using  Eqs.(\ref{13}),(\ref{15}) and (\ref{17}) the compact result
\begin{equation}\label{18}
\partial_\alpha {\cal
R}_\alpha =\mathbb{E}_\alpha\{\mathbb{Q}_\alpha {\cal K}'\}{\cal
R}_\alpha
\end{equation}
is found which directly leads to
\begin{equation}\label{19}
 \partial_{\alpha\alpha} G_\alpha =( {\cal
K}'|\mathbb{Q}_\alpha  {\cal K}')_\alpha
\end{equation}
and to the final expression for the Onsager term
\begin{equation}\label{20}
G^{(2)}=( {\cal K}'|\mathbb{Q}_0  {\cal K}')_0 \quad.
\end{equation}
Both the Mori product and the projector $\mathbb{Q}_0$ are related
to the Hamiltonian ${\cal K}_0$ of the free system. Thus $G^{(2)}$
can explicitly be calculated. According to Eqs.(\ref{5000}) the
conjugate variables $\nu^i_0$ enter in this expression that must
again be eliminated with Eqs.(\ref{6000}) to obtain the
$A_i$-dependence of the Onsager term $G^{(2)}$.

\subsection{Cumulants  for $\bm{n> 2}$ }

From the above treatment for the Onsager term it is obvious that
higher derivatives are needed to calculate the $ G^{( n )}$ for
general values of $n $. For this propose it is useful to
generalize the definitions of $ \mathbb{E}_\alpha $,
 of $ \mathbb{P}_\alpha $ and of the Mori product.

First a commutative product  $ {\cal  B }_1 \ast {\cal  B }_2\ast
\ldots \ast {\cal B}_n $ of an arbitrary  number  of  operators is
introduced \cite{comast}. The operation $ \mathbb{E}_\alpha $ on
such a $\ast$-product  is defined as a mapping to an (Hilbert
space) operator given by
\begin{eqnarray}\label{21}
&{}&\mathbb{E}_\alpha\;\{{\cal  B }_1 \ast {\cal  B }_2\ast \ldots
\ast {\cal B}_n\} = \\\nonumber &{}& \int_0^1 {\rm d}\, \lambda_1
\ldots \int_0^1 {\rm d} \lambda_n\;\mathbb{T}\big[{\cal
B}_1(\lambda_1){\cal B}_2(\lambda_2)\ldots{\cal
B}_n(\lambda_n)\big ] \quad.
\end{eqnarray}
The $\lambda_k$-dependence of the ${\cal B}_k(\lambda_k)$ is given
by Eq.(\ref{11}) and $ \mathbb{T}$ represents the thermodynamic
(or imaginary time) ordering operator. It orders the ${\cal
B}_k(\lambda_k)$ operators with increasing $\lambda_k$ from the
left to the right.

 Next the definition of the Mori product is
generalized by
\begin{equation}\label{211}
   ({\cal V}|{\cal  B }_1 \ast  \ldots \ast {\cal
B}_n )_\alpha=\langle {\cal
     V}^\dag\mathbb{E}_\alpha \{
    {\cal  B }_1 \ast  \ldots \ast {\cal
B}_n \}\rangle_\alpha\:.
\end{equation}
The bra must always  be an ordinary operator. It is just the ket
which can be an ordinary operator, a $\ast$-product or even linear
combinations of these objects. Assuming in Eq.(\ref{17}) that
${\cal U}$ represents such an object, the generalized projectors
$\mathbb{P}_\alpha$ and $\mathbb{Q}_\alpha$ are  still defined by
these equations.  The characteristic projector relation $
\mathbb{P}_\alpha \mathbb{P}_\alpha=\mathbb{P}_\alpha$ remains
valid which again is easy to proof.

The above generalizations imply new properties of the modified
quantities. Some of these properties are listed in Appendix
\ref{app0}. Two key properties,  the derivatives of
$\mathbb{E}_\alpha$ and $ \mathbb{P}_\alpha $, are calculated in
Appendix \ref{app1} and in Appendix \ref{app2}, respectively.
There we find
\begin{eqnarray}\label{23}
\partial_\alpha\;\mathbb{E}_\alpha \{{\cal  B }_1\ast \ldots
\ast{\cal B}_n \}{\cal R}_\alpha = \quad &{}&\\\nonumber
\mathbb{E}_\alpha \{{(\mathbb{Q}_\alpha{\cal K}')\ast\cal B}_1\ast
\ldots\ast {\cal B}_n \}{\cal R}_\alpha
 &+& \mathbb{E}_\alpha \{ \partial_\alpha\;{\cal  B }_1\ast \ldots\ast
 {\cal B}_n \}{\cal R}_\alpha \:,
\end{eqnarray}
where  the inner derivative $ \partial_\alpha \;{\cal B}_1\ast
\ldots \ast{\cal B}_n $ has be calculated by the usual chain rule.

Introducing the  shorthand notation for the $\ast$-product $
\textbf{B}= {\cal  B }_1 \ast {\cal  B }_2\ast \ldots \ast {\cal
B}_n $ Eq.(\ref{23}) and  the definition (\ref{211}) leads to the
derivative of the generalized Mori product
\begin{equation}\label{24}
\partial_\alpha({\cal  U }| \textbf{B})_\alpha =
({\cal  U }|(\mathbb{Q}_\alpha{\cal K}')\ast \textbf{B})_\alpha +
(\partial_\alpha{\cal  U }| \textbf{B})_\alpha +({\cal  U
}|\partial_\alpha \textbf{B})_\alpha\:.
\end{equation}
In  Appendix \ref{app2} it is shown that the derivative of
$\mathbb{P}_\alpha$ is given by
\begin{equation}\label{25}
\partial_\alpha\;\mathbb{P}_\alpha \textbf{B} =\mathbb{P}_\alpha
\;(\mathbb{Q}_\alpha {\cal K}')\ast(\mathbb{Q}_\alpha
\textbf{B})+\mathbb{P}_\alpha \;
\partial_\alpha \textbf{B}
 \:.
\end{equation}
Both differentiation rules (\ref{24}) and (\ref{25}) are essential
for the following. To this point it is assumed that $ \textbf{B}$
represents an arbitrary  $\ast$-product. An extension, however, to
linear combinations of such products is obvious. Assuming the
usual addition, multiplication and differentiation  rules for
these linear combinations, the above Eqs.(\ref{24}) and (\ref{25})
also hold for  linear combinations. Thus $ \textbf{B}$ represents
in general  linear combinations of $\ast$-products in these
equations.

With all these extensions we are well-equipped to calculate  the
higher derivatives of ${\cal R}_\alpha $  by repeated application
of the above rules . With the notation
\begin{equation}\label{27}
 {\cal F }_{1}=\mathbb{Q}_0{\cal K}' \quad \textrm{and} \:\
 {\cal F }_{n}=-\frac{1}{n!}\,\mathbb{P}_0 \textbf{F}^{(n)} \quad(\textrm{for}\:
 n\geq 2)
\end{equation}
we find
\begin{equation}\label{18aaaa}
(\partial_\alpha )^n{\cal R}_\alpha
=\mathbb{E}_\alpha\{\mathbb{Q}_\alpha \textbf{F}^{(n)}\}{\cal
R}_\alpha
\end{equation}
with
\begin{eqnarray}\label{280}
\textbf{F}^{(1)}&=&{\cal K}'\\\label{28}  \textbf{F}^{(2)}&=&{\cal
F}_1\ast{\cal F}_1={\cal F}_1^{\bm 2}
\\\label{29}
 \textbf{F}^{(3)}  &=&{\cal F}_1^{\bm 3}+6{\cal F}_1\ast{\cal F}_2
 \\\label{30}
 \textbf{F}^{(4)}  &=&{\cal F}_1^{\bm 4}+12 {\cal F}_1^{\bm 2}\ast{\cal F}_2
+12 {\cal F}_2^{\bm 2} +24{\cal F}_1\ast{\cal F}_3 \\\nonumber
\textbf{F}^{(5)} &=&{\cal F}_1^{\bm 5}+20 {\cal
F}_1^{\bm3}\ast{\cal F} _2 +60 {\cal F} _1\ast{\cal F}
_2^{\bm2}+60{\cal F}_1^{\bm2}\ast {\cal F} _3\\&{\quad}&\quad\:
    +120 {\cal F}_2 \ast{\cal F}  _3+120{\cal F} _1 \ast{\cal F} _4
\end{eqnarray}
for low $n$ values. Powers of the $\ast$ multiplication are
denoted by bold power exponents (compare Eqs.(\ref{28})).

 For
general values $n\geq1 $ the  $\textbf{F}^{(n+1)}$  are given by
\begin{equation}\label{31}
   \textbf{F}^{(n+1)}=\!\!\!\!\sum_{(k_1,\ldots,k_{n})}\!\!\!\!\!\!{}'
   \quad\frac{(n+1)!}
   {k_1!\;k_2! \ldots k_n! }\;{\cal F}_{1}^{\bm{k_1}}\ast
   {\cal F}_2^{\bm{k_2}}\ldots \ast{\cal F}_{n}^{\bm{k_n}}
\end{equation}
where the sum  runs over all $k_i= 0,1,2,\ldots$ with the
constrain
\begin{equation}\label{32}
\sum _{i=0}^n i\: k_i =n+1 \:.
\end{equation}
The general expression (\ref{31}) can be proofed by mathematical
induction.

With these results we obtain for the cumulants
\begin{equation}\label{26}
G^{(n+1)}=({\cal  K}'| \mathbb{Q}_0 \textbf{F}^{(n)})_0 \quad.
\end{equation}
Note that Eq.(\ref{27}) and Eq.(\ref{31}) permit a  recursive
determination of all the $\textbf{F}^{(n)}$. Consequently this
also applies to all cumulants $G^{(n)}$. Some examples are
\begin{eqnarray}\label{32a}
 &G^{(3)}&=({\cal K}'| \mathbb{Q}_0 (\mathbb{Q}_0
 {\cal K}')^{\textbf{2}})_0\\[3mm]\nonumber
& G^{(4)} & = \big( {
 \cal K}'\big|\mathbb{Q}_0
 \big\{(\mathbb{Q}_0 {\cal K}')^{\textbf{3}}
 -3(\mathbb{Q}_0 {\cal K}')\ast[\mathbb{P}_0
 (\mathbb{Q}_0 {\cal K}')^{\textbf{2}}]\big\}\big)_0\\\nonumber
 &=& \big( {
 \cal K}'\big|\mathbb{Q}_0
 \big\{-2(\mathbb{Q}_0 {\cal K}')^{\textbf{3}}
 +3(\mathbb{Q}_0 {\cal K}')\ast[\mathbb{Q}_0
 (\mathbb{Q}_0 {\cal K}')^{\textbf{2}}]\big\}\big)_0 \:
\end{eqnarray}
which show a  nested structure of the projectors. The different
forms of the $G^{(4)}$ result just using the
$\mathbb{P}_0+\mathbb{Q}_0=\bm{1}$. All the expressions are
treatable as they have to be calculated with respect by the bare
Hamiltonian $ {\cal K}_0 $. Again  the dummy variables $\nu^i_0 $
enter and have to be eliminated by use of Eq.(\ref{6000}). Thus
all terms  of the expansions for the Gibbs potential (\ref{5}) and
for the self energy (\ref{6}) can in principle be calculated.

Even for the classical system the representation  of the higher
cumulants  based on projectors seems to be adventurous  compared
to the approach based on generalized Maxwell relations
\cite{gyis}. Indeed relations like Eq.(\ref{31}) for general $n$
values have not been published for the latter approach.

In this context it should be added that Eq.(\ref{32a}) applied to
Ising systems  leads to the known  third and forth order terms for
these systems \cite{gyis,nata-hop}.(For an explicit check of this
claim the simplification (\ref{211cl}) of below can be used.)

\section{Discussion \label{secdis}}
\subsection{General remarks}
The expansion of the Gibbs potential (\ref{5}) and the expansion
of the self-energy (\ref{6}), together with the expression
(\ref{5001}) for the coefficient $G^{(1)}$ and the relations
(\ref{26}),(\ref{27}) and (\ref{31}), represent the most general
result of this work.

According to the formal derivation the perturbation $ {\cal K}'$
is completely arbitrary. Moreover no restrictions enter for the
bar Hamiltonian $ {\cal K}_0 $ or for the  operators of the
observation level. Thus one particle problems as well as
many-particles problems can be treated with the presented results.
In the most general case the observation level may   contain even
many-particles operators. The work of  Biroli and Cugliandolo
\cite{bicu} for the quantum version of the spherical p spin glass
model represents such an interesting approach.

 It is always possible to extend an observation level by adding
arbitrary operators $\widehat{{\cal A}}^j$ to the original set.
With the presumption that all the corresponding Lagrange
multipliers $\widehat{\nu}_\alpha^j $ satisfy the condition
$\widehat{\nu}_1^j=0$ the physical problem is not modified. Indeed
the Hamiltonian and consequently the exact Gibbs potential do not
change at all by such an extension. The projectors $\mathbb{P}_0$,
however, differ which leads to different terms of the expansions.
Thus different observation levels lead to different expansions.
This implies that the quality of approximations like truncations
of the power series depend on the chosen observation level.

The latter conclusion is clearly illustrated by the following
simple limiting case. Adding formally the interaction
$\widehat{{\cal A}}={\cal K}'$ with $\widehat{\nu}_1=0 $  to any
observation level Eq.(\ref{27}) and Eq.(\ref{26}) yields ${\cal
F}_1=0 $ and $ G^{(n\geq2)}=0 $ , respectively. This implies that
the first two terms of the expansion give the exact result whereas
any truncation of the original expansion represents an
approximation.

The present approach requires that the bare Hamiltonian can be
represented as linear combination of the elements of the
observation level \cite{comhubb}.  Nevertheless a reduction of the
observables of the observation level is  possible. This can simply
be achieved by transforming just a part of the parameters of the
bare Hamiltonian to conjugate variables. Again any reduction leads
to different expansions.

As far as no approximations are performed all observation levels
are equivalent. For specific problems this freedom can be used to
choose  a special observation level  which leads to good or fast
converging approximations. Certainly such a procedure requires in
general some physical intuition for the system under
consideration. Similar situations show up  in all the approaches
that are based on projector methods. In analogy it is generally
expected that symmetry breaking operators  should be included in
the observation level apart from the bare Hamiltonian. For two
particle interactions the complete set of all one particle
operators should be an appropriate  choice  for the observation
level (compare Sec. \ref{two}). Considering systems of Bosons or
Fermions in second quantization it should usually be sufficient to
span the observation level by  all the occupation number operators
(compare  Sec. \ref{secBose}).

\subsection{Classical systems  and the quasi-classical case}
For \textit{classical}  systems the $\ast$ product reduces to the
ordinary product. All quantities commute and  the Mori
product(\ref{211}) reduces  to
\begin{equation}\label{211cl}
   ({\cal V}|{\cal  B }_1 \ast{\cal
B}_2 \ldots  )_\alpha\longmapsto\langle {\cal
     V}^\dag
    {\cal  B }_1 {\cal  B }_2  \ldots  \rangle_\alpha\;,
\end{equation}
which usually simplifies the calculation.

For special quantum systems the situation may appear  that all
operators ${\cal A  }_i$ and the  interaction ${\cal K }'$ form a
set of commuting operators.  It is exclusively the algebra of this
set which enters in the $G^{(n)}$ and thus the replacement
(\ref{211cl}) can be used to calculate these quantities. In the
following we call such situations the \textit{quasi-classical
case}.

The approach for the fermionic spin glass model  by Rehker and
Oppermann \cite{opp} is such a quasi-classical case. Thus it is
obvious that the results of this fermionic system are very similar
to the results \cite{tap,I,II} for the SK spin glass model
\cite{sk}.   Note that for this system a quantum treatment becomes
necessary if a transverse magnetic field is added. The existing
approach, however, does not contain such a treatment.

\subsection{High-temperature expansion\label{sechte}}

The limiting case of an empty observation level is  of special
interest. No variable at all is Laplace  transformed. This implies
a vanishing $ {\cal K}_0$ and a quasi-classical situation where
Eq.(\ref{211cl}) can be employed. The thermodynamic potential
(\ref{4}) is nothing else as the free energy $\ln \, Z$ in
high-temperature approximation. The statistical operator
simplifies to
\begin{equation}\label{dmin1}
\widetilde{\cal R}_0 ={\bm 1} /\widetilde{Z}_0 \quad\textrm{with
\quad} \widetilde{Z}_0 =\textrm{Tr}\;{\bm 1}
\end{equation}
and the general projector reduces to $\widetilde{\mathbb{P}}_0$
which projects any ${\cal U}$
\begin{equation}\label{d1}
\widetilde{\mathbb{P}}_0 {\cal U}=\langle{\cal U}\rangle_0
\end{equation}
onto the $ 1$-direction. Setting
\begin{equation}\label{d2}
\ln Z  =  \sum_{n=0}^\infty\frac{ (\ln Z)^{(n)}}{n!}
\end{equation}
and
\begin{equation}\label{d3}
c_n= \big\langle\,({\cal  K }')^{\bm n}\,\big\rangle_0
\end{equation}
we find immediately
\begin{eqnarray}
  (\ln Z)^{(1)} &=& c_1 \\\nonumber
  (\ln Z)^{(2)} &=& c_2-c_1^2 \\\nonumber
  (\ln Z)^{(3)} &=& c_3 -3 c_2 c_1+2 c_1^3\\\nonumber
 (\ln Z)^{(4)} &=& c_4-4 c_3 c_1-3 c_2^2 +12 c_2 c_1^2 -6
 c_1^4\quad\ldots\quad.
\end{eqnarray}
These results agree with the  Ursell-Mayer expansion, a cumulant
expansion of the free energy for classical or quasi-classical
systems \cite{kubo} and demonstrates the close relation of the
present work to these former approaches.

\subsection{\label{two}Consequences for two particle interactions }

In the  typical many-body problem the interaction is a two
particle interaction. Focusing on this case the  thermodynamic
Hamiltonians is represented as
\begin{equation}\label{33}
{\cal K}_\alpha =    \sum_i \bm{\nu}^i_\alpha\cdot \bm{A}^{\,i} +
\frac{\alpha}{2 } \sum_{i,j}
\bm{A}^{\,i}\cdot\bm{C}^{ij}\,\bm{A}^{\,j} \quad
\textrm{with}\quad \bm{C}^{ii}=0\:.
\end{equation}
The indices $i$ and $j$ number the individual particles. For fixed
$i$ the components ${\cal A}^{i,m}$ and $\nu^{i,m}_\alpha $ of the
vector matrices $\bm{A}^i$ and $\bm{\nu}^i_\alpha$ represent one
particle operators and the corresponding Lagrange parameters,
respectively. The elements of the square matrices
$\bm{C}^{ij}=\bm{C}^{ji}$ describe the interaction. To include the
general case it is assumed that the observation level is spanned
by the set of all (linear independent) single particle operators
$\bm{A}^i$.

The factorization property of the expectation values with respect
to   ${\cal K}_0 $ leads to simplifications. The mean-field
contribution (\ref{5001}) reduces to
\begin{equation}\label{34}
G^{(1)}=\langle{\cal K}'\rangle_0=\frac{1}{2 } \sum_{i,j}
\langle\bm{A}\rangle^{\,i}\cdot\bm{C}^{ij}\,\langle\bm{A}\rangle^{\,j}
\:.
\end{equation}
Using Eq.(\ref{17}) we find
\begin{equation}\label{35}
 {\cal F }_{1}
    = \frac{1}{2}\sum_{i,j}  \widetilde{\bm{ A}}^{i}\cdot\bm{C}^{ij}
    \widetilde{\bm{
    A}}^{j}\quad \textrm{with}
    \quad \widetilde{\bm{ A}}^{i}=\bm{ A}^{i}-
    \langle\bm{ A}^{i}\rangle\:,
\end{equation}
which leads to the Onsager term
\begin{equation}\label{36}
 G^{(2)}
    = \frac{1}{2} \sum_{i,j}  \:(\widetilde{\bm{ A}}^{i}
    \cdot\bm{C}^{ij}\widetilde{\bm{
    A}}^{j}|\widetilde{\bm{ A}}^{i}\cdot\bm{C}^{ij}\widetilde{\bm{
    A}}^{j})_0\:.
\end{equation}
The latter result implies several interesting features. First of
all the Onsager term is a superposition of the correlation
functions which do not factorize in the general quantum case $
(i\neq j)$
\begin{eqnarray}
\label{37}
 &{}&(\widetilde{{\cal A}}^{\,i,m_i}
    \:\widetilde{{\cal
    A}}^{\,j,m_j}|\widetilde{{\cal A}}^{\,i,m_i'}\widetilde{{\cal
    A}}^{\,j,m_j'})_0  \\\nonumber
&{}&\quad\quad\quad\quad\quad\quad\quad\neq (\widetilde{{\cal
A}}^{\,i,m_i}
    |\widetilde{{\cal A}}^{\,i,m_i'})_0\;(\widetilde{{\cal
    A}}^{\,j,m_j}|\widetilde{{\cal
    A}}^{\,j,m_j'})_0\:.
\end{eqnarray}
In contrast, these correlation functions factorize for classical
systems or for the quasi-classical case. The presence of these
\textit{quantum fluctuations} can be the origin of essential
differences between classical and quantum systems (for an example
compare below).

Thermal averaging gives finite contributions to the Onsager term
(\ref{36}) only for such terms where each $\widetilde{{\cal
A}}^{\,i,m_i}$ has at least one partner $\widetilde{{\cal
A}}^{\,i,m_i'}$. Thus the sums  are double sums and triple sums do
not appear in Eq.(\ref{36}). In contrast to this behavior, the
second order term of expansions of the free energy $\ln Z$ leads
to triple sums as this term is given by $(1|\widetilde{{\cal
K}}'\ast\widetilde{{\cal K}}')_0$, according to \cite{kubo}.

 A similar behavior holds for the higher cumulants. In any order the
number of terms of the free energy expansion extends the number of
terms of the Gibbs potential approach where at maximum $n$-fold
sums arise. This conclusion is some indication that the
correlations are more efficiently treated by the Gibbs Potential
expansion. Note that in particular these arguments apply for
systems with long ranged interactions.

In this context it is remarked that a diagrammatic interpretation
of the expansion can be given completely analog to the classical
systems \cite{gyis,papot}. As pointed out by these authors, the
weak point of this method is that the vertex weight and the
combinatorial factors can not be calculated systematically and
thus the Feynman rules are not known. Nevertheless, some
conclusions are possible from these diagrammatic approaches. In
particular, and relevant for this work, it is found that all
diagrams are connected. Note that this can also be concluded from
the general cumulant theory \cite{kubo}.

The last two conclusions cause the well known fact that for
infinite-ranged models the expansions truncate for both  non
random and random interactions in the thermodynamic limit.

Consider first the case that  the matrix elements of $\bm{C}^{ij}$
scale as $N^{-1}$ and are non random. From the discussion above
follows all terms $G^{(n)}$ with $n\geq 2 $ are sub-extensive and
can be neglected for large $N$. Thus the expansions Eq.(\ref{5})
and Eq.(\ref{6})  reduce to the usual mean-field expressions
\begin{eqnarray}\label{38}
 G\big( \langle\bm{A}\rangle^i\big)  &=& S_0
 \big( \langle\bm{A}\rangle^i\big)+\frac{1}{2 } \sum_{i,j}
\langle\bm{A}\rangle^{\,i}\cdot\bm{C}^{ij}\,\langle\bm{A}
\rangle^{\,j}
\\\label{38a}
  \bm{\nu}^i_1 &=& -\frac{\partial G}{\partial \langle\bm{A}\rangle^{\,i}}
  \quad \quad(\textrm{for} \quad \bm{C}^{ij}\sim N^{-1})\:.
\end{eqnarray}
Recall that the function $S_0$ is the  entropy of the
non-interacting system. Thus the Eqs.(\ref{38}) and (\ref{38a})
completely determine  all thermodynamic properties for the given
parameters $\bm{\nu}^i_1$ and $\bm{C}^{ij}$ of the Hamiltonian
(\ref{33}).

Finally we consider random, infinite ranged systems   where the
matrix elements of $\bm{C}^{ij}$ are independent random variables,
or where the $\bm{C}^{ij}$ are proportional to a random variables.
For these cases the scaling $\bm{C}^{ij}\sim N^{-1/2}$ has to be
used to get the right $N$ dependence for the extensive quantities.
All cumulants with $n\geq3$ are sub-extensive and it is just the
Onsager (\ref{36}) that must be added to Eq.(\ref{38}). For
classical systems these conclusions are well known and for quantum
systems they are in agreement with \cite{bicu}.

\section{Quantum spin glass ${\bm(s=\frac{1}{2} \label{secspin} )}$}

\subsection{General non isotropic case \label{secspin1}}

A system of  $N$ quantum spins $ \bm{s}_{\,i}$ (with $s={1\over2}$
and $\hbar=1$) is considered in the presence of external fields $
\bm{h}_i$. The spins interact via an infinite ranged spin-spin
interaction
 $J_{ij}  $ and are described by the  Hamiltonian
\begin{equation}\label{s1}
{\cal H} =   - \sum_i \bm{h}_i\cdot \bm{s}_{\,i} - \frac{1}{2 }
\sum_{i,j}J_{ij}\; \bm{s}_{\,i}\cdot\bm{\Gamma}\,\bm{s}_{\,j}\:,
\end{equation}
where the dot denotes the scalar product in the three dimensional
real space. The bonds $ J_{ij}= J_{ji}$ ( with $J_{ii}=0$) are
independent random variables with zero means and standard
deviations $ J N^{-1/2}$. We consider a general spin-spin
interaction and $\bm{\Gamma}$ represents an arbitrary symmetric
tensor  with real eigenvalues $\gamma^\mu$ and $\mu=x,y,z $
\cite{ww}. The norm of $\bm{\Gamma}$ is denoted by
\begin{equation}\label{s01}
\gamma^2=(\gamma^x)^2+(\gamma^y)^2
+(\gamma^z)^2=\textrm{tr}\bm{\Gamma}^2\;.
\end{equation}

 The complete set of all one particle
operators $ s_i^\mu$ with $i=1,\ldots N$ and $\mu=x,y,z$ is used
as observation level. Setting
$\bm{m}_i=\langle\bm{s}_{\,i}\rangle_\alpha=\langle\bm{s}_{\,i}\rangle_1$
and $m_i=|\bm{m}_i| $ the entropy of the non-interacting system as
function of the $\bm{m}_i$ is well known and given by
\begin{equation}\label{s2}
S_0=-\sum_i (\frac{1}{2}+m_i)\ln(\frac{1}{2}+m_i)
+(\frac{1}{2}-m_i)\ln(\frac{1}{2}-m_i)\:.
\end{equation}
The operator
\begin{equation}\label{s3}
    {\cal K}_0 = \sum_i \bm{\nu}_i\cdot \bm{s}_{i}
\end{equation} governs the calculation of the
expectation values  and the  Mori products. Simplifying the
notation the variables $\nu^i_0$ and $\nu^i_1$ of Sec.\ref{s2a}
are denoted by $\bm{\nu}_i$ and by $ \beta \bm{h}_i $ ,
respectively. From Eq.(\ref{6000}) we find
\begin{equation}\label{s4}
{\nu}_i=  \;2\,\textrm{artanh}\,( 2 m_i)\textrm{\quad and }\quad
\frac{\bm{\nu}_i}{\nu_i}=\frac{\bm{m}_i}{m_i} =\bm{e}_i \:.
\end{equation}
 Note that these  equations have to be used to eliminate
 the dummy variables $\bm{\nu}_i$.
 Eq.(\ref{6}) leads to the thermic equation of states
\begin{equation}\label{s5}
\bm{e}_i \;2\,\textrm{artanh}\,( 2 m_i)=\beta (\bm{h}_i-
\bm{\Sigma}_i)\:.
\end{equation}
According to Sec.(\ref{two})   all cumulants $G^{(n)}$ with $n\geq
3 $ can be neglected in the thermodynamic limit. Thus the internal
fields $-\bm{\Sigma}_i$ are given by
\begin{equation}\label{s6}
\beta \bm{\Sigma}_i =-
\frac{\partial}{\partial\bm{m_i}}\Big(G^{(1)}+\frac{G^{(2)}}{2}\Big)\:.
\end{equation}

To calculate the terms $G^{(1)}$ and $G^{(2)}$  we note that with
Eq.(\ref{s1})
\begin{equation}\label{s7}
{\cal K}' =  \,  \frac{\beta}{2 } \,\sum_{i,j}J_{ij}\;
\bm{s}_{\,i}\cdot\bm{\Gamma}\,\bm{s}_{\,j}
\end{equation}
holds. All expectation values factorize, the mean-field
contribution (\ref{5001}) becomes
\begin{equation}\label{s8}
 G^{(1)} = \,\frac{\beta}{2 } \,\sum_{i,j}J_{ij}\;
\bm{m}_{\,i}\cdot\bm{\Gamma}\,\bm{m}_{\,j}\;
\end{equation}
and from the definition (\ref{17})
\begin{equation}\label{s7a}
\mathbb{Q}_0\,{\cal K}' =   \, \frac{\beta}{2 }\,
\sum_{i,j}J_{ij}\;
\widetilde{\bm{s}}_{\,i}\cdot\bm{\Gamma}\,\widetilde{\bm{s}}_{\,j}
\quad
\textrm{with}\quad\widetilde{\bm{s}}_{\,i}=\bm{s}_{\,i}-\bm{m}_{i}
\end{equation}
results.

The calculation of the Onsager term is straightforward but needs
some more effort.  Again using the factorization property and
Eq.(\ref{a123}) the  term (\ref{20}) takes the form
\begin{equation}\label{s7b}
G^{(2)}=  \, \frac{\beta^2}{2 }\, \sum_{i,j}J^2_{ij}\,X_{ij}
\end{equation}
with
\begin{equation}\label{s7cc}
  X_{ij} =X_{ji}=\big(\widetilde{\bm{s}}_{\,j}\cdot\bm{\Gamma}\,
\widetilde{\bm{s}}_{\,i}\,\big|\widetilde{\bm{s}}_{\,i}\cdot\bm{\Gamma}\,
\widetilde{\bm{s}}_{\,j}\big)_0\:.
\end{equation}
This Mori product is treated  in  Appendix \ref{app3}. The result
is split in longitudinal, transverse and mixed contributions
\begin{equation}\label{s20}
X_{ij}= X_{ij}^{LL}+X_{ij}^{LT}+X_{ji}^{LT}+X_{ij}^{TT}\:,
\end{equation}
which are given by
\begin{eqnarray}\label{s21}
X_{ij}^{LL} &=&\frac{1}{\nu_i'}\frac{1}{\nu_j'}\:
\Gamma_{ij}^{2}\\\label{s22}
 X_{ij}^{LT}  &=&
 \frac{1}{\nu_i'}\;\frac{m_j}{\nu_j}\:
 \big\{(\bm{\Gamma}^2)_{ii}-\Gamma_{ij}^{2}\big\}\\\nonumber
X_{ij}^{TT}&=&\frac{1 }{8}\big\{\frac{m_i+m_j}{\nu_i+\nu_j} -
\frac{m_i-m_j}{\nu_i-\nu_j}\big\}\:\textrm{tr}\,(\bm{e_i}\times)\,\bm{\Gamma}
\,(\bm{e_j}\times)\,\bm{\Gamma}
\\\nonumber
 &+&\:\frac{1 }{8}\big\{ \frac{m_i+m_j}{\nu_i+\nu_j} +
\frac{m_i-m_j}{\nu_i-\nu_j}\big\}\\\label{s23}&{}& \,\big \{
\gamma^2\,-(\bm{\Gamma}^2)_{ii}
 -(\bm{\Gamma}^2)_{jj}+\Gamma_{ij}^{2}\big\} \:.
\end{eqnarray}
The antisymmetric tensor associated with the cross product is
denoted by
 $ (\bm{e}_i \mathbf{\times})$.
The quantities $ \Gamma_{ij}$ and $(\bm{\Gamma}^2)_{ii}$ are
components
\begin{equation}\label{s24corr}
\Gamma_{ij}=\bm{e_i}\cdot\bm{\Gamma}\bm{e_j}\quad;\quad
(\bm{\Gamma}^2)_{ii}=\bm{e}_i\cdot\bm{\Gamma}^2\bm{e}_i
\end{equation}
of the tensors $\bm{\Gamma}$ and $\bm{\Gamma}^2 $. To write the
Eqs.(\ref{s22}) and (\ref{s23}) as short as possible we have not
completely eliminated the dummy variables $\nu_i$ which, however,
can easily be done with Eq.(\ref{s4}) and with
\begin{equation}\label{s16at}
\frac{1}{\nu_i'}=\frac{\partial
m_i}{\partial\nu_i}=\Big(\frac{1}{4}-m_i^2\Big)
 \quad.
\end{equation}
Note that the quantities $1/\nu_i'$ and $m_i/{\nu_i}$ have a
physical meaning. They represent the longitudinal and the
transverse susceptibilities of the bare system.

Putting things together we  find finally with the usual
replacement $J_{ij}^2\rightarrow J^2/N$
\begin{equation}\label{s24}
G= S_0 + \,\frac{\beta}{2 } \,\sum_{i,j}J_{ij}\;
\bm{m}_{\,i}\cdot\bm{\Gamma}\,\bm{m}_{\,j}\,+\,\frac{\beta^2J^2}{4
N}\,\sum_{i j}X_{ij}
\end{equation} and
\begin{equation}\label{s25}
\beta\bm{h}_i =\bm{\nu_i}-\beta\sum_j J_{ij}\,\bm{\Gamma m}_j
-\,\frac{\beta^2J^2}{2 N}\,\sum_{j} \frac{\partial
X_{ij}}{\partial \bm{m_i}}\:.
\end{equation}
For given external fields $\bm{h}_i$ the magnetizations $\bm{m_i}$
are determined  by the equation of states (\ref{s25}). Provided
that these solutions $\bm{m}_i$ are explicitly known all other
thermodynamic properties follow from the Gibbs potential
(\ref{s24}).  Recall that these equation are exact in the
thermodynamic limit. They are equivalent to the TAP free energy
and the TAP equations for the SK model\cite{tap,I}.

To the best of the authors' knowledge such results for the quantum
$s=1/2$ spin  system have not been published previously. From the
analogy to the TAP equations it is expected that these equations
will shed some light on the spin glass problem in quantum systems.
Certainly the present results are just the basis for this purpose
and  additional work is needed in this direction.

Some elementary aspects are presented in the following subsections
skipping  points which are interesting from the spin glass point
of view.  Such questions need additional efforts which are far
beyond the scope of this work.

Before we go into  details  a general aspect of our results  is
pointed out. Note that the transverse contributions (\ref{s23}) to
the Onsager term exhibit \textit{ energy denominators}. Such
denominators  are a characteristic feature of all expansions for
quantum systems. The denominators are absent in the longitudinal
and the mixed contributions given by Eq.(\ref{s21}) and
(\ref{s22}) , respectively. As all operators commute this is
obvious for the longitudinal part. For the mixed contributions it
is a consequence of the cyclic property of the trace operation.
\subsection{Comparison with classical models}
Let us next work out the differences to the  classical spin glass.
For this we consider a system described again by Eq.(\ref{s1})
where  $ \bm{S}_{\,i}^\textrm{cl}$  represent  classical vector
spins in three dimensions of length $S_{\,i}^\textrm{cl}=1/2 $.
Such a treatment leads  to modified Eqs.(\ref{s24}) and
(\ref{s25}). Obviously  both the entropy term and  the $\nu_i$
have  to be replaced by the classical expressions for
$S_0^{\textrm{cl}}$ and for the $\nu_i^{cl}(m_i)$.

As quantum fluctuations are absent the Onsager term simplifies
(compare Eqs.(\ref{37})). We obtain
\begin{equation}\label{s256a}
  X_{ij}^\textrm{cl} =\big\langle(\widetilde{\bm{S}}_{\,j}^\textrm{cl}
  \bm{\cdot\Gamma}
\widetilde{\bm{S}}_{\,i}^\textrm{cl})\:(\widetilde{\bm{S}}_{\,i}^\textrm{cl}
\bm{\cdot\Gamma}\,
\widetilde{\bm{S}}_{\,j}^\textrm{cl})\big\rangle_0=
\textrm{tr}\:\bm{\chi}_i^\textrm{cl}\,\bm{\Gamma}\bm{\chi}_j^\textrm{cl}\,
\bm{\Gamma}\:,
\end{equation}
where the local susceptibility tensor $\bm{\chi}_i^\textrm{cl}$ is
given by
\begin{equation}\label{256b}
    \left(\chi^{\textrm{cl}}_i\right)^{\mu \bar{\mu}}=\left\langle
    (\widetilde{{S}}_{\,i}^{\textrm{cl}})^\mu\:
(\widetilde{{S}}_{\,i}^{\textrm{cl}})^{\bar{\mu}}\right\rangle_0
\:.
\end{equation}
Writing again $X_{ij}^\textrm{cl}$ as sum of  longitudinal,
transverse and  mixed terms we find for the transverse term
\begin{equation}\label{s26}
    \Big(X_{ij}^{TT}\Big)^\textrm{cl}=\frac{m_i m_j}{\nu_i^\textrm{cl} \nu_j^\textrm{cl}}\big \{
\gamma^2-(\bm{\Gamma}^2)_{ii}
 -(\bm{\Gamma}^2)_{jj}+\Gamma_{ij}^{2}\big\}\;,
\end{equation}
whereas the other contributions to $X_{ij}^\textrm{cl}$ are still
given by Eqs.(\ref{s21}) and (\ref{s22}) with $\nu_i(m_i)$
replaced by $\nu_i^{cl}(m_i)$.

Note that no explicit representation exists for $\nu_i^{cl}(m_i)$.
Nevertheless this function is well defined via the inverse
function
\begin{equation}\label{sc27}
   m_i= \frac{1}{2} \coth
   \left(\frac{\nu_i^\textrm{cl}}{2}\right)-\frac{1}{\nu_i^\textrm{cl}}
\end{equation}
and can be represented by the power expansion
\begin{equation}\label{sc28}
  \nu_i^\textrm{cl}=  12 m_i+\frac{144 m_i^3}{5}+\frac{19008
   m_i^5}{175}+O\left(m_i^7\right)\;.
\end{equation}
It should be remarked that for the isotropic case $
\bm{\Gamma}=\bm{1} $ all the results for the classical spins are
in agreement with the previous work \cite{bmclass,vul} where in
addition the  explicit expressions for the classical entropy
function can be found.

\subsection{Paramagnetic phase for isotropic interactions\label{secpara}}
To keep the discussion as simple as possible   we specialize to
the isotropic case and set $ \bm{\Gamma}=\bm{1} $. This implies
$\gamma^2=3$. To explore the  paramagnetic limit $m_i\rightarrow 0
$ all terms have to be  expanded. We find
\begin{equation}\label{s2500}
   \nu_i=4 m_i+\frac{16
   }{3}m_i^3+O\left(m_i^5\right)
\end{equation}
and
\begin{equation}\label{s251}
   X_{ij } = \frac{3}{16}-\frac{1}{12}
 (5 m_i^2
 +5 m_j^2 + 2\bm{m}_i\bm\cdot\bm{ m}_j )+O\left(m_i^4\right)\:.\:
\end{equation}
\

These expansions lead to the Gibbs potential
\begin{equation}\label{s2511}
G= N\ln 2 + N \frac{3}{64}\, \beta^2 J^2  \quad \textrm{for}\quad
m_i=0
\end{equation}
and to the equations of states
\begin{eqnarray}
\nonumber\label{s25112}
  \beta \bm{h}_i&=& 4  \bm{m}_i -\beta\sum_j
  J_{ij}\;\bm{m}_j\\
 &+&\frac{\beta^2J^2}{12}\big(5\bm{m}_i+
N^{-1} \sum_j \bm{m}_j\big)
 + O\left(m_i^3\right)\;.
\end{eqnarray}

Obviously the paramagnetic state (all $\bm{m_i}=0 $) is a solution
of the Eqs.(\ref{s25112}) for vanishing fields $\bm{h_i}=0 $.
Employing the standard relations $ U =-
\partial G/\partial \beta $ and $ S=G+\beta U$ gives the internal energy $U $ and the
entropy $ S $  of this state
\begin{equation}\label{s025}
U= - N \frac{3}{32}\, \beta J^2 \quad \textrm{and} \quad S=N\ln 2
- N \frac{3}{64}\, \beta^2 J^2\:,
\end{equation}
 respectively. For high  temperatures $\beta\rightarrow 0$ these
results are expected from the high-temperatures expansion
according to Sec.\ref{sechte}. In the low temperature regime,
however, both the internal energy and the entropy are not
acceptable as both quantities diverge to the negative infinity.
Consequently there must be a phase with solutions  $\bm{m_i}\neq 0
$ for low temperatures in the zero field case.

For  further analysis we focus on the  singularities of the
susceptibility matrix $\chi_{ij}^{\mu \bar{\mu}}= \partial
m_i^\mu/\partial h_j^{\bar{\mu}} $. The inverse matrix is
determined by
\begin{equation}\label{sssqw}
(\chi^{-1}_{ij})^{\mu\bar{\mu}}=\partial h_i^\mu/\partial
m_j^{\bar{\mu}}=-\beta^{-1}\partial^2 G/\partial m_i^{\mu}\partial
m_j^{\bar{\mu}}\:.
\end{equation}
Reintroducing the expansion parameter $\alpha$ we find from
Eq.(\ref{s25112})
\begin{equation}\label{s25113}
 \beta (\chi^{-1}_{ij})^{\mu \bar{\mu}} =I_{ij}\:\delta^{\mu
 \bar{\mu}}\;\;\;\textrm{with}\;\; I_{ij}^\mu
 =a\delta_{ij}
-b \frac{J_{ij}}{J}\; +\frac{c}{N}
\end{equation}
and with the coefficients
\begin{equation}\label{s25114}
a= 4 +\frac{5\,(\alpha \beta J)^2}{12}\quad;\quad b=\alpha\beta
J\quad;\quad
   c =\frac{(\alpha \beta J)^2}{12}\;.
\end{equation}

From random-matrix theory \cite{random} the eigenvalue spectrum of
the matrix $I_{ij}$ is well known. It is a superposition of a
continuous part
\begin{equation}\label{spec1}
\lambda(x)=a - b x \quad \textrm{with} \quad -2\leq x\leq2
\end{equation}
and  one discrete  eigenvalue
\begin{equation}\label{spec2}
\lambda_0 =a+c +\frac{b^2}{c}
\end{equation}
which may be isolated from $\lambda(x)$. For the special values of
the expansion parameter $\bar{\alpha}(x)$ and $\bar{\alpha}_0$ the
eigenvalues $\lambda(x)$ and $\lambda_0$ vanish , respectively.
These values  and their absolute values are calculated to
\begin{equation}\label{spec3}
\bar{\alpha}(x)=  2 \,\frac{3 x\,\pm i\sqrt{\,60-9
   x^2}}{5\beta J} \quad;\quad
\bar{\alpha}_0=\pm\, i \,\frac{4\sqrt{2}}{\beta
   J}
\end{equation}
and to
\begin{equation}\label{spec4}
|\bar{\alpha}(x)|^2=\frac{48}{5}\:(\beta
   J)^{-2}\quad \;\quad |\bar{\alpha}_0|^2=
\,{32}\:(\beta
   J)^{-2}\;,
\end{equation}
respectively. Realize  that $\bar{\alpha}(x)$ has a finite
imaginary part for the possible $x$-values $|x|\leq 2$ . Thus
vanishing eigenvalues of the inverse of the susceptibility   and
singularities for the susceptibility are only possible for the
complex $\bar{\alpha}$ values. In the complex plane the
singularities $\bar{\alpha}(x)$ are located on two sectors of the
circle with the radius $|\bar{\alpha}(x)|$. As no intersections of
these sectors with real axis exist there are no singularities for
real values of $\alpha$.

These singularities are of high importance, as already pointed out
in \cite{I}, to which the reader is referred for more details .
The term-by-term treatment of the approach of Sec.(\ref{sgen}) can
only be justified in the region of the \textit{complex $\alpha
$-plane } in which the power expansion is convergent. The
convergence criterium for a Taylor expansion is given by
$|\alpha|<\rho$ where $\rho$ is the radius of convergence. The
distance from the origin ($\alpha=0$) to the nearest singular
point determines this radius $\rho$ . Thus, in the present case,
$\rho=|\bar{\alpha}(x)|$ holds \cite{diff} and the expansion for
the paramagnetic solution can only be justified  for temperatures
$T$ above the critical temperature
\begin{equation}\label{spec5}
T_c=\sqrt{5/3}\;J/4 \;.
\end{equation}
Below $T_c$ the expansion and consequently the paramagnetic
solution breaks down and at least  one new phase must exist.

For the SK model and the classical vector spin glass (compare
below) the singularities of the susceptibility in the complex
$\alpha$ plane show a similar behavior. For these systems the
singularities are located \textit{everywhere} on a circle in the
complex $\alpha$ plane. The circle intersects the positive and the
negative real axis,  which implies two real values of $\alpha$.
From this point of view both the quantum spin glass and the
classical  models are quite similar. An important difference,
however, results. For the classical models a physical staggered
susceptibility diverges at $T_c$. Such a behavior is not found for
the quantum spin glass.

To complete  the analysis of the paramagnetic phase we calculate
the local susceptibility
\begin{equation}\label{spec6}
\chi_{\textrm{loc}}=\frac{1}{N}\sum_i\chi_{ii}^{\mu\mu}=\beta\int_{-2}^2
\frac{\varrho(x)\;\textrm{d}x }{ 4 -\beta J x +5/12( \beta J)^2
}\;
\end{equation}
 of the quantum spin glass. Employing the
Wigner semi- circle law \cite{random} for the density
$\varrho(x)=\sqrt{(4-x^2)}/(2\pi)$  of the eigenvalues and
introducing the reduced temperature
\begin{equation}\label{}
t=T/T_c=4\sqrt{3/5}\;T/J
\end{equation}
we find by an integration
\begin{equation}\label{spec7}
\chi_{\textrm{loc}}^2=\frac{5 (t^4+1)+4
   t^2-\left(t^2+1\right) \sqrt{25 t^4-10
   t^2+25}}{6\;J^2 \;t^2}\:,
\end{equation}
which holds for $t>1$.  The function $\chi_{\textrm{loc}}(t)$
exhibits a maximum at the critical temperature $t=1$ and decreases
with increasing temperature  from the  value $J
\chi_{\textrm{loc}}(1)=\{\frac{1}{3} \left(7-2
   \sqrt{10}\right)\}^{1/2} =0.474498 $ .  We find
   $\chi_{\textrm{loc}}\rightarrow\beta/4$ for $\beta\rightarrow
   0$, in agreement with the direct high-temperature expansion.

The Eqs.(\ref{s25112}) and (\ref{spec7}) give a complete
thermodynamic description of the paramagnetic phase above the
critical temperature given by Eq.(\ref{spec5}). Again the author
is not aware of any  work which has claimed these results before.

It is of some interest to compare with the classical vector spin
glass. The classical results are well known \cite{bmclass} but can
easily be rederived. We find with Eq.(\ref{sc28}) by expansion
\begin{eqnarray}\label{spec8corr}
\nonumber
 X_{ij }^{\textrm{cl}} = \frac{1}{48}-\frac{1}{12}
 \Big\{m_i^2 + m_j^2\
   \Big\} +O \left(m_i^4\right)\;.
\end{eqnarray}
In contrast to   Eq.(\ref{s251}), terms proportional to $ \bm{m}_i
\cdot \bm{m}_j$  are not found. This results from the differences
between  $(X_{ij }^{TT})^{\textrm{cl}}$ and  $X_{ij }^{TT}$.

 The remaining calculation is completely analog to the
quantum case and leads for the Gibbs potential to
\begin{equation}\label{spec8}
G^{\textrm{cl}} = const. + N(\beta J)^2/192 \;.
\end{equation}
The matrix  $I_{ij}^{\textrm{cl}}$ which governs the singularities
 is calculated to
\begin{equation}\label{spec9}
(I_{ij}^\mu)^{\textrm{cl}}
 =(12 +\beta^2\alpha^2 J^2/12)\;\delta_{ij}
- \beta\alpha J_{ij}\;.
\end{equation}
The singular $\alpha$-values  are  located on a circle with radius
$ |\alpha|=12(\beta J)^{-1}$ . This leads to a  transition
temperature of
\begin{equation}\label{spec10}
   T_c^{\textrm{cl}}=J/12
\end{equation}
and to a local susceptibility of
\begin{equation}\label{spec11}
   \chi_{\textrm{loc}}^{\textrm{cl}}=\beta/12
\end{equation}
for $T\geq T_c^{\textrm{cl}}$  . The results for all quantities
differ.  These differences are exclusively caused by the different
equation of states of the bare system given by Eq.(\ref{s4}) and
by Eq.(\ref{sc28}). The different expressions for $ X_{ij}^{TT}$
have no influence.

Obviously these conclusions are restricted to the paramagnetic
phase at zero external fields. Incidentally we remark in this
context that the 'quantum' term $c/N$ of Eq.(\ref{s25113}) is
important for systems with an additional (infinite range)
ferromagnetic interaction $ J_0/N $. Indeed in this case an
additional contribution $-J_0/N$ has to be added to
Eq.(\ref{s25113}) leading to a competition with the term $c/N$ due
to the different signs.

The replica approaches \cite{bmquant,parra} claim for the critical
temperature of the quantum spin glass  $T_{c}\approx
J/(4\sqrt{3})$. These values are determined from the condition
$1=J\chi_{\textrm{loc}}(T_c)$ which is a consequence of the
assumption \cite{bmquant} that near $T_c$ the system has a
continuous behavior, like a second order phase transition. Such a
behavior, however, is not confirmed by the present work. Moreover
Eq.(\ref{s025}) leads for a temperature of $J/(4\sqrt{3})$ to a
negative entropy value of $S= -1.55685$, which is impossible for a
quantum system. Thus the  replica theory values for $T_c$ must be
rejected.

We conclude by noting that the discussion presented in this
subjection can be extended to the general  non-isotropic case and
to the low temperature phase on bases of the presented results.
For this propose all the tools \cite{II,complex}, developed
 for the understanding of the SK model on bases of the TAP
equations, can be transferred to the quantum spin glass. Work in
this direction for  the quantum SK model in presence of a
transversal magnetic field will be published separately
\cite{tobe}.

\section{The weakly non-ideal Bose gas \label{secBose}}

\subsection{The Gibbs potential in second order\label{secBose1}}
The model is described by the Hamiltonian
\[
{\cal H} = \sum_{\vk} \e_{\vk} {\cal N}_{\vk} +
\frac{U}{2V}\sum_{\vp \vq \vk} \, {b}_{\vp+\vk}^\dag
{b}_{\vq-\vk}^\dag {b}_{\vp} {b}_{\vq} \, \, ,
\]
where $\e_{\vk}=k^2/2m$ is the free-gas spectrum, the ${b}_{\vk}$,
${b}_{\vk}^\dag$ are bosonic operators,  the ${\cal N}_{\vk}={
b}_{\vk}^\dag {b}_{\vk}$ are occupation number operators and $V$
is the volume of the gas. The interaction strength $ U$  is
assumed to be momentum-independent which corresponds  a  delta
function interaction  in real space.

We choose the set of all operators ${\cal N}_{\vk}$ as observation
level. Thus the operators ${\cal K}_0$, ${\cal K}_1$ and ${\cal
K}'$ introduced in Sec.\ref{sgen} are given by
\begin{equation}\label{b2}
{\cal K}_0=\sum_{\vk} \nu_{\vk} {\cal N}_{\vk}\quad,\quad {\cal
K}_1= \sum_{\vk}\b(\mu-\e_\vk){\cal N}_{\vk}
\end{equation}
and by
\begin{equation}\label{b2int}
{\cal K}'=-\frac{\b U}{2V}\sum_{\vp \vq \vk} \, {b}_{\vp+\vk}^\dag
{b}_{\vq-\vk}^\dag {b}_{\vp} {b}_{\vq}\;.
\end{equation}
 Again we simplify the notation.  The variables $\nu^i_0$ and
$\nu^i_1$ of Sec.\ref{s2a} are denoted by ${\nu}_\vk $ and by $
\b(\mu-\e_\vk) $ respectively. We work with the second
quantization. Thus  $ G_\alpha $ represents the Legendre
transformation of the logarithm of the grand-canonical  partition
function and $\mu$ is the chemical potential.

Let us introduce the notation $ n_\vk=\langle {\cal
N}_{\vk}\rangle_\alpha $. Then the entropy $ S_0$ of the
noninteracting system as function of these variables $n_\vk$  is
needed and explicitly given by
\begin{equation}\label{b3}
S_0   =\sum_{\vk}\{\: (n_{\vk}+1)\ln (n_{\vk}+1)-n_{\vk}\ln
n_{\vk}\}\;.
\end{equation}
Employing Eq.(\ref{6000}), this leads to
\begin{equation}\label{b4}
\nu_\vk=\ln n_{\vk}-\ln (n_{\vk}+1)
\end{equation}
which has to be used to eliminate the dummy variables $\nu_\vk$.
Note that an   equivalent form of the latter  equation
\begin{equation}\label{b5}
n_{\vk}=\{ \exp(-\nu_\vk)-1\}^{-1}
\end{equation}
is the Bose function.

The mean-field or Hartree Fock contributions are given by
Eq.(\ref{5001}) and  are immediately  calculated  to
\begin{equation}\label{b6}
G^{(1)}=-\frac{\b U}{V}\left(\sum_\vp n_\vp\right)^2\;,
\end{equation}
where the relation (\ref{a100}) is used. Similar to the spin glass
system the treatment of the Onsager term is more complicated and
therefore is presented in some detail in Appendix \ref{app4}.
Using the result Eq.(\ref{a130}) together with Eq.(\ref{b3}) and
Eq.(\ref{b6}) we obtain the expansion of the Gibbs potential up to
second order in $U$
\begin{eqnarray}
 \nonumber
G(\beta,n_\vk)& = &\sum_{\vk}\{\: (n_{\vk}+1)\ln
(n_{\vk}+1)-n_{\vk}\ln n_{\vk}\}\;\\\nonumber &-&\frac{\b
U}{V}\sum_{\vp\vq} n_\vp n_\vq
 \\\nonumber
  &+&\frac{\beta^2 U^2}{ V^2}\sum_{\vp\:\vq \:
 \vk}
\frac{n_{\vp+\vk}\,n_{\vq-\vk}\:(1 +\,n_{\vp}+\,n_{\vq})}
{\nu_{\vp+\vk} +\nu_{\vq-\vk}-\nu_{\vp}-\nu_{\vq}}\\\label{b7}
&+&O\left(U^3\right)\;,
\end{eqnarray}
where in principle all the $ \nu_\vk$ in the denominator of the
second order term have to be replaced according to relation
(\ref{b4}).

The general Eqs.(\ref{6000}) and (\ref{6}) lead immediately  to
the equation of states
\begin{equation}\label{b8}
\b(\e_\vk-\mu)= \ln\frac{n_\vk+1}{n_\vk}-\b\Sigma_\vk^{(1)}
-\b\Sigma_\vk^{(2)}-\b\widetilde{\Sigma}_\vk^{(2)}+O
\left(U^3\right)
\end{equation}
where $ \Sigma_\vk^{(1)}$ and
$\Sigma_\vk^{(2)}+\widetilde{\Sigma}_\vk^{(2)}$ represent the
first and the second order contributions to the self energy given
by
\begin{equation}\label{b9}
\Sigma_\vk^{(1)} = \frac{2U}{V}\, \sum_{\vp } n_\vp\;,
\end{equation}
by
\begin{equation}\label{b10}
\Sigma_\vk^{(2)}= \frac{2\beta U^2}{ V^2}\sum_{\vp\:\vq }
  \frac{n_{\vp+\vq}n_{\vk-\vq}-n_\vp(1+  n_{\vp+\vq}+n_{\vk-\vq})}
  {\nu_{\vk} +\nu_{\vp}-\nu_{\vp+\vq}-\nu_{\vk-\vq}}
\end{equation}
and by
\begin{eqnarray}\label{b11}
 &\widetilde{\Sigma}_\vk^{(2)}&=\frac{2\beta U^2}
  { V^2 }\sum_{\vp\:\vq }
\frac{1 }
  {
  (\nu_{\vk} +\nu_{\vp}-\nu_{\vp+\vq}-\nu_{\vk-\vq})^2 }
  \\[3mm]\nonumber
&{}&\!\!\!\!\!\!\!\!\!\left\{\frac{ n_\vp(1+ n_{\vp+\vq})(1+
n_{\vk- \vq})}{(1+ n_{\vk})} -\frac{(1 +n_\vp)
 n_{\vp+\vq}n_{\vk-\vq}}
  {n_\vk}\right\}\:.
\end{eqnarray}
Note that  $\widetilde{\Sigma}_\vk^{(2)}$ results from the
application of the chain rule for differentiation to
$(\nu_{\vp+\vk} +\nu_{\vq-\vk}-\nu_{\vp}-\nu_{\vq})^{-1}$.

The main part of the general results Eqs.(\ref{b7})-(\ref{b11})
are well known in literature.  Trivially this applies to the  zero
order contributions. All first order terms represent the usual
Hartree-Fock expressions \cite{haque,baym} for an interaction that
is momentum-independent. The second order contribution
$\Sigma_\vk^{(2)}$ to the  self energy looks like  typical
expressions calculated  by the Green's functions approaches
\cite{second}.

Contributions like $\widetilde{\Sigma}_\vk^{(2)}$ that contain the
square of $(\nu_{\vk} +\nu_{\vp}-\nu_{\vp+\vq}-\nu_{\vk-\vq})$ in
the denominator are usually not considered or discussed by the
standard Green's function treatments. In this situation some
arguments are presented  for the relevance of this contribution in
the next subsection.
\subsection{Discussion}
The first argument for the importance of the contribution
$\widetilde{\Sigma}_\vp^{(2)}$ is rather general. Note that this
quantity enters in the susceptibility matrix  that is defined as $
\chi_{\vp \vq}=-\partial n_\vp/\partial(\epsilon_\vq-\mu)$. Indeed
the inverse matrix is given by
\begin{eqnarray}\label{b12}
  \chi_{\vp \vq}^{-1}&=&
-\frac{1}{\beta}\frac{\partial^2 G}{\partial n_\vp
\partial n_\vp }\\\nonumber
   &=&\frac{\delta_{\vp \vq}}{\beta n_\vp(1+n_\vp)} +\frac{2 U}{V}
+\frac{\partial \Sigma_\vp^{(2)}}{\partial
n_\vq}+\frac{\partial\widetilde{\Sigma}_\vp^{(2)}}{\partial
n_\vq}+O \left(U^3\right)\:.
\end{eqnarray}
This matrix governs the stability of the system and a fundamental
property of this matrix is the symmetry relation $ \chi_{\vp
\vq}^{-1}= \chi_{\vq \vp}^{-1}$. Thus neglecting or modifying some
terms of $\chi_{\vq \vp}^{-1}$ results in general  to a violation
of this relation with serious  consequences.

Various investigations exist for the weakly interacting Bose gas
in literature. In particular the equation of state at the critical
point of the Bose-Einstein transition was recently discussed in
 \cite{baym}. As argued in this  work, the
transition
\begin{equation}\label{b13}
n_\vq\gg 1 \quad
\end{equation}
can be used in all equations. With this approximation we find in
leading order
\begin{equation}\label{b14}
\Sigma_\vk^{(2)}=-\frac{2\beta U^2}{ V^2}\sum_{\vp\:\vq }
\frac{n_{\vp+\vq}n_{\vk-\vq}n_\vp\left(\frac{n_{\vk}}{n_{\vp}}
-\frac{n_{\vk}}{n_{\vp+\vq}}
-\frac{n_{\vk}}{n_{\vk-\vq}}\right)}{1
+\frac{n_{\vk}}{n_{\vp}}-\frac{n_{\vk}}{n_{\vp+\vq}}
-\frac{n_{\vk}}{n_{\vk-\vq}}}
\end{equation}
and
\begin{equation}\label{b15}
\widetilde{\Sigma}_\vk^{(2)}=-\frac{2\beta U^2}{
V^2}\sum_{\vp\:\vq }\frac{ n_{\vp+\vq}n_{\vk-\vq}n_\vp} {1
+\frac{n_{\vk}}{n_{\vp}}-\frac{n_{\vk}}{n_{\vp+\vq}}
-\frac{n_{\vk}}{n_{\vk-\vq}}}\quad.\:\:\quad
\end{equation}
Both contributions $\Sigma_\vk^{(2)}$  and
$\widetilde{\Sigma}_\vk^{(2)}$ to the second order self energy are
of the same order of magnitude, which demonstrates again the
relevance of $\widetilde{\Sigma}_\vk^{(2)}$.

Using $n_\vq\gg 1 $ we find as approximation for the equation of
states
\begin{equation}\label{b16}
\e_\vk-\mu= \frac{1}{\b n_\vk}-\frac{2 U}{V}\, \sum_{\vp }
n_\vp+\frac{2\beta U^2}{ V^2}\sum_{\vp\:\vq }
n_{\vp+\vq}n_{\vk-\vq}n_\vp \;,
\end{equation}
and  as approximation for the inverse of the susceptibility
\begin{equation}\label{b17}
\chi^{-1}_{\vk \vk'}= \frac{\delta_{\vk \vk'}}{\b n_\vk^2}+\frac{2
U}{V}-\frac{2\beta U^2}{ V^2}\sum_{\vq } n_{\vq+\vk}(2\,
n_{\vq+\vk'}+ n_{\vk'-\vq})\;.
\end{equation}
By a change of the summation index it is elementary to show the
symmetry of $\chi^{-1}_{\vk \vk'}$ , which is not surprising as
both $\Sigma_\vk^{(2)}$ and $\widetilde{\Sigma}_\vk^{(2)}$ are
included in Eq.(\ref{b17}).

All perturbations  of $\chi^{-1}_{\vk \vk'}$ are of the order
$V^{-1}$ and can therefore be neglected \cite{care}. This implies
that  it is  the $\vk=0$ mode which becomes instable at the
critical temperature. Such a behavior was assumed in the analysis
\cite{baym} without any proof and the present approach confirms
this assumption.

Considering next the equation of states (\ref{b16}) we  note that
$n_\vq\gg 1 $ implies the approximation (compare Eq.(\ref{b5}))
\begin{equation}\label{b18}
n_\vq^{-1}=-\nu_\vq = \beta (\hat{\e}_\vq -\hat{\mu})
\end{equation}
where we have introduced in addition the 'dressed' energy $
\hat{\e}_\vq $ and the 'dressed' chemical potential $\hat{\mu}$ as
the $\vq$-dependent and the $\vq$-independent part of
$-\nu_\vq/\b$ , respectively. Setting the average value of the
density $n=V^{-1}\sum_\vp n_\vp$ the equation of states
(\ref{b16}) can be written in terms of the 'dressed' variables
\begin{eqnarray}
\e_\vk-\mu =&\hat{\e}_\vk&-\hat{\mu}-2\, U n
\quad\quad\quad\quad\\\nonumber
  &+& \frac{2 U^2}{\beta^2
V^2}\sum_{\vp\:\vq }\frac{1}{(\hat{\e}_{\vp+\vq}-\hat{\mu})
(\hat{\e}_{\vk-\vq}-\hat{\mu})(\hat{\e}_{\vp}-\hat{\mu})}\:.
\end{eqnarray}
This is in complete agreement to the equation derived in
\cite{baym}, both within the frameworks of Green's function and of
Ursell operators. Thus we conclude that the  expansion of the
Gibbs potential is  an alternative to other approaches for
many-body systems of identical particles.

Apart from this important conclusion for the present work  we
remark that  detailed numerical investigations of the presented
result are expected to be an interesting object of further
research. Indeed, a complete numerical analysis of the
Eqs.(\ref{b7}) and (\ref{b8}), which should include the entire
temperature regime,  may potentially give some new insight to the
Bose-Einstein condensation for weakly interacting gases. This
expecting  is based on the impressive success of these equations
for calculating the shift of the critical temperature of the
transition \cite{baym,haque}.

\section{Conclusions \label{seccon}}
In this paper we have presented a systematic power expansion of
the Gibbs potential for arbitrary many-particle systems including
in particular all kinds of quantum systems.  Developing and
employing  generalized projector techniques we were able to
present   explicit formulas which permit the calculation of the
contributions  up to an arbitrary order of the expansion for
general  systems. After a detailed discussion of the general
results the method is applied to two non-trivial systems, the
quantum spin glass  with infinite ranged interactions and the
weakly interacting Bose gas. The contributions up  to the Onsager
terms, which are the lowest beyond mean-field terms, have been
worked out leading to new results or confirming recent results for
these special systems.

The present method has several advantages compared to other
techniques. The method can be applied for all kinds of interacting
systems including, in particular, systems of identical particles,
classical or quantum spin systems and combinations of these
systems. No other technique seems to have such a wide spectrum for
applications.

Compared to the efforts needed within the Green's functions or the
Ursell operator approach the expansion of Green's is rather
simple, direct and straightforward. In particular no partial
summations are needed for the present approach to find the
mean-field and the beyond mean-field contributions to the self
energy.

As a further advantage the present approach usually gives criteria
directly for the convergence of the expansion. Within the
framework of other techniques additional investigations are
usually needed to obtain this information.

The application of the cavity method \cite{mpv} is very common for
spin glasses and related problems. This method often allows
convincing interpretations of the low order terms of  formal
expansions. The cavity approach was originally developed for
classical systems. To treat in addition  quantum spin glasses work
\cite{qsg} has been presented which uses Trotter-Suzuki
transformations and maps the quantum spin systems into classical
spin models. Such  treatments work only for special problems and
can not be generalized to all quantum systems. Thus the existing
extensions of the cavity method to quantum systems are restricted.
The present Gibbs potential approach, however,  works for general
quantum systems.

Summing up we conclude that the power expansion approach may
potentially represent a serious alternative, to the other,
well-settled methods to treat the statics of many-particle
systems. Certainly  more applications must be worked out to
confirm this possibility.

\begin{acknowledgments}
The author would like to thank K.W. Becker, B. Drossel, N. Grewe,
H. Sauermann and O. Zobay for interesting discussions.
\end{acknowledgments}

\appendix*
\section{}
\subsection{\label{app0} Some properties  of the Mori scalar product}
Since the projector formalism is not very common, we list some
elementary relations of this approach that are used in the present
work. For further details and explicit proofs of these relations
we refer to \cite{fs}.

Apart from the general properties which are required for any
scalar product the definition of the original Mori product
(\ref{401}) implies some additional properties. Let ${\cal U}$ and
${\cal V}$ be two arbitrary elements, then the relations
\begin{equation}\label{aaa1}
( {\cal U}|{\cal V})_{\alpha}=( {\cal V}^\dag|{\cal
U^\dag})_{\alpha}=( {\cal V}|{\cal U})_{\alpha}^*=( {\cal
U^\dag}|{\cal V^\dag})_{\alpha}^*
\end{equation}
and the Kubo Identity
\begin{equation}\label{aaa2} ( {\cal U}|[{\cal
K}_\alpha,{\cal V}])_{\alpha}=\langle[{\cal V},{\cal
U^\dag}]\rangle_{\alpha}
\end{equation}
can be proofed from the definition and the invariance of the trace
to cyclic permutations. The property (\ref{aaa2}) implies the
useful relation for special operators ${\cal W}$. If $[{\cal
K}_\alpha,{\cal W}]= \omega {\cal W} $ with $\omega\neq 0$ is
satisfied
\begin{equation}\label{0a0}
\omega ( {\cal V}|{\cal W})_{\alpha}= \langle{[{\cal W},\cal
V}^\dag]\rangle_{\alpha}
\end{equation}
results.

The Eqs.(\ref{401}) and (\ref{aaa1})  yield for the scalar product
of the unit operator $1$ with an observable
\begin{equation}\label{aaa3}
( 1|{\cal U})_{\alpha}= \langle{\cal U}\rangle_{\alpha}\quad
\textrm{and}\quad ( {\cal U}|1 )_{\alpha}= \langle{\cal U^
\dag}\rangle_{\alpha}\:,
\end{equation} and for fluctuations of observables
$ \widetilde{{\cal U}}= {\cal U}-\langle{\cal U}\rangle_{\alpha}$
one  finds
\begin{equation}\label{aaa4}
( \cal U|\widetilde{{\cal V}})_{\alpha}=( {\cal
U}|\widetilde{{\cal V}})_{\alpha}=( \widetilde{\cal
U}|\widetilde{{\cal V}})_{\alpha}=( \cal U|{\cal V})_{\alpha}-
\langle{\cal U^\dag}\rangle_{\alpha}\langle{\cal
V}\rangle_{\alpha}\:.
\end{equation}

The projectors $\mathbb{P}_\alpha$ and $\mathbb{Q}_\alpha$ are
Hermitian in  Liouville space and idempotent which implies with
the definitions (\ref{17}) the relations
\begin{equation}\label{a123xx}
({\cal U}| \mathbb{P}_\alpha {\cal V })_\alpha =(\mathbb{P}_\alpha
{\cal U}| {\cal V })_\alpha=(\mathbb{P}_\alpha {\cal
U}|\mathbb{P}_\alpha {\cal V })_\alpha
\end{equation}
and
\begin{equation}\label{a123}
({\cal U}| \mathbb{Q}_\alpha {\cal V })_\alpha =(\mathbb{Q}_\alpha
{\cal U}| {\cal V })_\alpha=(\mathbb{Q}_\alpha {\cal
U}|\mathbb{Q}_\alpha {\cal V })_\alpha\:.
\end{equation}
For later use it is noted that
\begin{equation}\label{a0000}
(\widetilde{{\cal A}}^k| \mathbb{Q}_\alpha {\cal U} )_\alpha=0
\end{equation}
according to Eq.(\ref{a123}) and $\mathbb{Q}_\alpha\widetilde{\cal
A}^k=0$.

Let c be a complex number and let  $\textbf{B}= {\cal  B }_1 \ast
{\cal B }_2\ast \ldots \ast {\cal B}_n $ be an $\ast$-product.
Then the definition (\ref{211}) immediately yields
\begin{equation}\label{a000}
({\cal U}| \textbf{B}\ast c\; 1)_\alpha= c ({\cal U}| \textbf{B}
)_\alpha\:.
\end{equation}
Again for later use  we finally note that
\begin{equation}\label{a04}
\big(1| (\mathbb{Q}_\alpha {\cal K}')\ast(\mathbb{P}_\alpha
\textbf{B})\big)_\alpha=0
\end{equation}
holds. To prove this result we recall that $\mathbb{P}_\alpha
\textbf{B}$ represents a linear combination of the unite operator
and the $\widetilde{\cal A}^k$, whereas $\mathbb{Q}_\alpha {\cal
K}'$ does not contain such terms. Thus the thermal averaging
eliminates all contributions.

\subsection{\label{app1} The derivative of $ \bm{\mathbb{E}_\alpha}$}
Let ${\cal X}$ be an ordered  product of $n$ operators ${\cal
B}_k(\lambda_k)$
\begin{equation}\label{a1}
{\cal X}= \mathbb{T} \:{\cal B}_1 (\lambda_1){\cal B}_2
(\lambda_2)\ldots {\cal B}_n (\lambda_n)
\end{equation}
where the $\lambda$ dependencies  are given by Eq.(\ref{11}).
Assuming $ \lambda_1<\lambda_2< \ldots \lambda_n $ this product is
already ordered  and can be rewritten as
\begin{equation}\label{a2}
{\cal X}= e^{\lambda_1 {\cal K}_{\alpha}}{\cal B}_1
e^{(\lambda_2-\lambda_1 ) {\cal K}_{\alpha}} {\cal B}_2
e^{(\lambda_3-\lambda_2 ) {\cal K}_{\alpha}}  \ldots {\cal B}_n
e^{-\lambda_n  {\cal K}_{\alpha}}\:.
\end{equation}
From Eq.(\ref{12}) one finds by elementary substitutions
\begin{equation}\label{a3}
\partial_\alpha e^{(\lambda_{k+1}-\lambda_k ){\cal K}_{\alpha}}=
e^{-\lambda_k {\cal
K}_{\alpha}}\int_{\lambda_k}^{\lambda_{k+1}}\frac{\partial{\cal
K}_{\alpha}}{\partial\alpha}\big(\lambda\big)\, \textrm{d} \lambda
\:\: e^{\lambda_{k +1} {\cal K}_{\alpha}}
\end{equation}
and obtains
\begin{eqnarray*}
\partial_\alpha{\cal X}&=& \int_{0}^{\lambda_{1}}\frac{\partial{\cal
K}_{\alpha}}{\partial\alpha}\big(\lambda\big)\, \textrm{d} \lambda
\:{\cal X}\\  &+& {\cal
B}_1(\lambda_1)\int_{\lambda_{1}}^{\lambda_{2}}\frac{\partial{\cal
K}_{\alpha}}{\partial\alpha}\big(\lambda\big)\, \textrm{d}
\lambda\:  {\cal B}_2 (\lambda_2)\ldots {\cal B}_n
(\lambda_n)\\&\vdots&\\&+&
    {\cal
B}_1(\lambda_1){\cal B}_2 (\lambda_2)\ldots
\int_{\lambda_{n-1}}^{\lambda_{n}}\frac{\partial{\cal
K}_{\alpha}}{\partial\alpha}\big(\lambda\big)\, \textrm{d}
\lambda\: {\cal B}_n (\lambda_n)\\&+&  {\cal
X}\:\int_{\lambda_{n}}^{0}\frac{\partial{\cal
K}_{\alpha}}{\partial\alpha}\big(\lambda\big)\, \textrm{d}
\lambda\:,
\end{eqnarray*}
provided that  all the operators ${\cal B}_k $ are independent of
$\alpha$. The integral of the last term is rewritten with
Eq.(\ref{11}) as
\begin{equation}\label{a4}
\int_{\lambda_{n}}^{0}\frac{\partial{\cal
K}_{\alpha}}{\partial\alpha}\big(\lambda\big)\, \textrm{d}
\lambda\:=\int_{\lambda_{n}}^{1}\frac{\partial{\cal
K}_{\alpha}}{\partial\alpha}\big(\lambda\big)\, \textrm{d}
\lambda\: - \:\mathbb{E}_\alpha \{\partial_\alpha{\cal K}_{\alpha}
\}\:.
\end{equation}
Recalling the definition of ordering operator $\mathbb{T}$ the
expression for $\partial_\alpha{\cal X}$ simplifies to
\begin{equation}\label{a5}
\partial_\alpha{\cal X}=-\:{\cal X}\mathbb{E}_\alpha \{\partial_\alpha{\cal
K}_{\alpha} \}\:+\:\int_{0}^{1}\textrm{d}
\lambda\:\mathbb{T}\,\frac{\partial{\cal
K}_{\alpha}}{\partial\alpha}(\lambda)\,{\cal X}\:.
\end{equation}
From Eqs.(\ref{12},\ref{15},\ref{17})
\begin{eqnarray}
\nonumber
\partial_\alpha{\cal K}_{\alpha} &=& {\cal K}' +\sum_i {\cal A}^{i}
\partial_\alpha\nu_\alpha^i \\\nonumber
  &=& \mathbb{Q}_\alpha{\cal K}' +\langle{\cal
K}'\rangle_\alpha + \sum_i \langle{\cal
A}^{i}\rangle_\alpha\partial_\alpha\nu_\alpha^i\\
\nonumber&=&\mathbb{Q}_\alpha{\cal K}'+c
\end{eqnarray}
results, where $c$ is a  number. For such numbers  the relations
$\mathbb{E}_\alpha\{c\}=c $ and $\mathbb{T}c {\cal X}= c {\cal X}$
hold. Replacing $\partial_\alpha{\cal K}_{\alpha}$ by
$\mathbb{Q}_\alpha{\cal K}'+c$ in Eq.(\ref{a5}) leads to
\begin{equation}\label{a6}
  \partial_\alpha{\cal X}=-\:{\cal X}\mathbb{E}_\alpha \{ \mathbb{Q}_\alpha{\cal K}'\}\:+\:\int_{0}^{1}\textrm{d}
\lambda\:\mathbb{T}(\mathbb{Q}_\alpha{\cal K}')(\lambda)\,{\cal X}
\end{equation}
as the terms proportional $c$ cancel.

 Note that the result (\ref{a6}) does not
 change for any other order of the operators $ B_k$ in
 Eq.(\ref{a1}).
 Therefore the above restriction on the $\lambda_k$ can be dropped
 and the $ \lambda_k $-integrations yields finally
\begin{eqnarray}
 \nonumber
&\,&  \partial_\alpha \mathbb{E}_\alpha  \{{\cal B }_1\ast  \ldots
\ast{\cal B}_n \} = \mathbb{E}_\alpha \{\:{\mathbb{Q}_\alpha{\cal
K}'\ast\cal B}_1\ast \ldots\ast {\cal B}_n \}\\\label{a7} &\,&
\quad -\mathbb{E}_\alpha \{\:{\cal B }_1\ast  \ldots \star{\cal
B}_n \}\;\mathbb{E}_\alpha \{\:\mathbb{Q}_\alpha {\cal K}'\}\quad.
\end{eqnarray}
With Eq.(\ref{18}) this result is equivalent to Eq.(\ref{23})  in
the text provided that all operators are independent of $\alpha$.
If the ${\cal B}_k $ depend on $\alpha$ it is obvious that the
inner derivatives have to be added and one gets the full
Eq.(\ref{23}).
\subsection{\label{app2} The derivative of $ \bm{\mathbb{P}_\alpha}$}
 Let $\textbf{B}= {\cal  B }_1 \ast {\cal  B }_2\ast \ldots \ast {\cal
B}_n $ be an $\ast$-product. Then the
 definition (\ref{17}) immediately yields
\begin{eqnarray}\label{a8}
&&\partial_\alpha\:\mathbb{P}_\alpha \textbf{B} =
\partial_\alpha\:(1|\textbf{B})_\alpha\quad \quad\quad\quad
\\[3mm]\nonumber
 &\quad &+ \sum_{ij}\, \widetilde{{\cal
A}}^i\,\Gamma_\alpha^{ij}\:\partial_\alpha\,( \widetilde{{\cal
A}}^j| \textbf{B})_\alpha \,+\,\sum_{ij}\, \widetilde{{\cal
A}}^i\,\big[
\partial_\alpha\:\Gamma_\alpha^{ij}\big]( \widetilde{{\cal
A}}^j|\textbf{B} )_\alpha \:.
\end{eqnarray}
From Eq.(\ref{24}) one finds
\begin{equation}\label{a9}
\partial_\alpha\:(1|\textbf{B})_\alpha= (1|
(\mathbb{Q}_\alpha{\cal K}')\ast\textbf{B})_\alpha +(1|
\partial_\alpha\textbf{B})_\alpha
\end{equation}
and
\begin{equation}\label{a10}
\partial_\alpha\:(\widetilde{{\cal
A}}^i|\textbf{B})_\alpha= (\widetilde{{\cal A}}^i|
(\mathbb{Q}_\alpha{\cal K}')\ast\textbf{B})_\alpha
+(\widetilde{{\cal A}}^i|
\partial_\alpha\textbf{B})_\alpha\:.
\end{equation}
Again with Eq.(\ref{17}) these two relations
 permit to rewrite the first two terms of
Eq.(\ref{a8}) which leads to
\begin{eqnarray}\label{a11}
\partial_\alpha\:\mathbb{P}_\alpha \textbf{B}& =& \mathbb{P}_\alpha
(\mathbb{Q}_\alpha {\cal K}')\ast\textbf{B} +\mathbb{P}_\alpha\:
\partial_\alpha \;\textbf{B}
\\[3mm]\nonumber
&\quad &+\sum_{ij}\, \widetilde{{\cal A}}^i\,\big[
\partial_\alpha\:\Gamma_\alpha^{ij}\big]( \widetilde{{\cal
A}}^j|\textbf{B} )_\alpha \:.
\end{eqnarray}
Differentiation of Eq.(\ref{16}) and employing again Eq.(\ref{24})
leads to
\begin{equation}\label{a12}
\partial_\alpha\:\Gamma_\alpha^{ij}=-\sum_{kl} \,
\Gamma^{ik}_\alpha \; ( \widetilde{{\cal A}}^k |\mathbb{Q}_\alpha
{\cal K}'\ast \widetilde{{\cal A}}^l)_\alpha \;
\Gamma^{lj}_\alpha.
\end{equation}
and with Eq.(\ref{17}) to
\begin{eqnarray}\label{a13}
&{}&\sum_j\big[\partial_\alpha\:\Gamma_\alpha^{ij}\big](
\widetilde{{\cal A}}^j|\textbf{B})_\alpha=- \sum_{k} \,
\Gamma^{ik}_\alpha \nonumber \big( \widetilde{{\cal A}}^k\big|
(\mathbb{Q}_\alpha {\cal
K}')\ast(\mathbb{P}_\alpha \textbf{B}) \big)_\alpha\\
& +& \sum_{k} \, \Gamma^{ik}_\alpha \big(\widetilde{{\cal
A}}^k\big| (\mathbb{Q}_\alpha {\cal K}')\ast(1|\textbf{B})_\alpha
\big)_\alpha\:.
\end{eqnarray}
The Mori product in the second term of Eq.(\ref{a13}) can be
written as $
 (1|\textbf{B})_\alpha\;(\widetilde{{\cal A}}^k| \mathbb{Q}_\alpha
{\cal K}' )_\alpha$ according to Eq.(\ref{a000}). Due  to
Eq.(\ref{a0000}) $(\widetilde{{\cal A}}^k| \mathbb{Q}_\alpha {\cal
K}' )_\alpha=0 $ holds and it is just the first term of
Eq.(\ref{a13}) which remains. The multiplication Eq.(\ref{a13})
with $\widetilde{{\cal A}}^i $ and a summation yields
\begin{equation}\label{a14}
\sum_{ij}\widetilde{{\cal
A}}^i\big[\partial_\alpha\:\Gamma_\alpha^{ij}\big](
\widetilde{{\cal A}}^j|\textbf{B} )_\alpha=-\mathbb{P}_\alpha
(\mathbb{Q}_\alpha {\cal K}')\ast(\mathbb{P}_\alpha \textbf{B})\:,
\end{equation}
where in addition the relation (\ref{a04}) was used.
Eq.(\ref{a14}) combined with Eq.(\ref{a11}) finally leads to
Eq.(\ref{25}) of the text.
\subsection{\label{app3} Calculation of
$ {\bm X_{ij}=\big(\widetilde{\bm{s}}_{\,j}\cdot\bm{\Gamma}\,
\widetilde{\bm{s}}_{\,i}\,\big|\widetilde{\bm{s}}_{\,i}\cdot\bm{\Gamma}\,
\widetilde{\bm{s}}_{\,j}\big)_0}$}

According to the definition (\ref{401}) of the Mori product we
have to calculate the quantity
\begin{equation}\label{s7c}
     X_{ij} =\int_0^1\,\textrm{d}\lambda
\Big\langle\big\{\widetilde{\bm{s}}_{\,j}\cdot\bm{\Gamma}\,
\widetilde{\bm{s}}_{\,i}\big\}\:\widetilde{\bm{s}}_{\,i}(\lambda)\cdot\bm{\Gamma}\,
\widetilde{\bm{s}}_{\,j}(\lambda)\Big\rangle_0 \:.
\end{equation}
First the  $ \lambda$-dependence of $\bm{s}_{i}(\lambda)$ is
considered. We set $ \alpha=0$ in the definition (\ref{11}) and
find from the  rules of the spin $s= {1\over2}$ algebra
\begin{equation}\label{s9}
\bm{s}_{i}(\lambda)= e^{\lambda \bm{\nu}_i \cdot
\bm{s}_{i}}\,\bm{s}_{i}\,e^{-\lambda \bm{\nu}_i \cdot \bm{s}_{i}}=
\bm{\Omega}_i(\lambda)\bm{s}_i\;,
\end{equation}
where the tensor $\bm{\Omega}_i(\lambda)$ describes a rotation of
the imaginary angle $-i\,\lambda \nu_i$ about the axis $\bm{e}_i$
\begin{eqnarray}\label{s10}
\bm{\Omega}_i(\lambda) &=& e^{-i \,\lambda\,(\bm{\nu}_i
\mathbf{\times})}=\bm{\Pi}_i
+\bm{\Omega}_i^T(\lambda)\quad\quad\quad\quad\quad\quad\quad\\[3mm]
\nonumber
 \bm{\Omega}_i^T(\lambda)  &=&- i \sinh(\lambda \nu_i)(\bm{e}_i
\mathbf{\times})-\cosh(\lambda \nu_i)(\bm{e}_i \mathbf{\times})^2
\end{eqnarray}
and where $\bm{\Pi}_i$ and $\bm{\Omega}_i^T(\lambda)$ are the
longitudinal and the transverse parts of $\bm{\Omega}_i(\lambda)$
, respectively . The tensor $\bm{\Pi}_i$  is the projector onto
the $\bm{e}_i$-direction and $ (\bm{e}_i \mathbf{\times})$
represents the antisymmetric tensor associated with the cross
product of two vectors. For further use the relations
\begin{equation}\label{s11}
\bm{\Pi}_i=\bm{\Pi}_i ^2 \quad \textrm{;}\quad (\bm{e}_i
\mathbf{\times})^2  = \bm{\Pi}_i-\bm{1} \quad \textrm{;}\quad
\bm{\Omega}_i^T(\lambda)\bm{\Pi}_i=0\quad;
\end{equation}
and
\begin{eqnarray}
\nonumber  \bm{\Omega}_i^T(\lambda_1+\lambda_2)&=&
 \bm{\Omega}_i^T(\lambda_1)\bm{\Omega}_i^T(\lambda_2)
 \quad;\quad\bm{\Omega}_i(\lambda)\bm{m}_i  =\bm{m}_i\quad;\\[3mm]\label{s11a}
\{\bm{\Omega}_i(\lambda)\bm{a}\}
\cdot\bm{b}&=&\bm{a}\cdot\,\bm{\Omega}_i(-\lambda) \bm{b}
\end{eqnarray}
are noted where $\bm{a}$ and $\bm{b}$ are arbitrary  vectors.
Using these relations and Eq.(\ref{s9}) the  $ \lambda$-dependent
part of Eq.(\ref{s7c}) is rewritten as
\begin{equation}\label{s13}
\widetilde{\bm{s}}_{\,i}(\lambda)\cdot\bm{\Gamma}\,
\widetilde{\bm{s}}_{\,j}(\lambda)=\big\{\bm{\Omega}_j(-\lambda)
\bm{\Gamma}\,\bm{\Omega}_i(\lambda)\;\widetilde{\bm{s}}_{\,i}\big\}
\cdot\widetilde{\bm{s}}_{\,j} \:.
\end{equation}

Let $ \bm{a}$ and $ \bm{b}$ be any two vectors (or  two vector
operators which commute with $\bm{s}_{\,i}$ ). From the well known
identity
\begin{equation}\label{s14}
({\bm{s}}_{\,i}\cdot{\bm{a}})(\bm{b}\cdot{\bm{s}}_{\,i})= \bm{a}
\cdot \Big\{ \frac{1}{4}-\frac{i}{2}
(\bm{s}_{\,i}\times)\Big\}\,\bm{b}
\end{equation}
it is elementary to prove the relation
\begin{equation}\label{s15}
\langle(\widetilde{\bm{s}}_{\,i}\cdot{\bm{a}})(\bm{b}\cdot
\widetilde{\bm{s}}_{\,i})\rangle_0=\bm{a}\cdot\
\bm{\Theta}_i(1/2)\,\bm{b}
\end{equation}
where we have introduced
\begin{equation}\label{s16}
 \bm{\Theta}_i(\lambda)=\;\frac{\bm{\Pi}_i}{\nu_i'}
+\frac{\bm{\Omega}_i^T(\lambda)}{ 4 \cosh (\nu_i/2)}
\end{equation}
and
\begin{equation}\label{s16a}
\frac{1}{\nu_i'}=\frac{\partial
m_i}{\partial\nu_i}=\Big(\frac{1}{4}-m_i^2\Big)
 \quad.
\end{equation}

Using   the partial result(\ref{s13}), the relations (\ref{s11})
and (\ref{s11a}) and applying the identity (\ref{s9}) twice we
find
\begin{equation}\label{s17}
X_{ij} =\int_0^1\,\textrm{d}\lambda\; \textrm{tr}\:
\bm{\Theta}_i({1}/{2}-\lambda)\,\bm{\Gamma}\,
\bm{\Theta}_j(\lambda-{1}/{2})\,\bm{\Gamma}
\end{equation}
where tr stands for the trace in the tree dimensional real vector
space. The $\lambda $ dependence of $\bm{\Theta}_i(\lambda)$ is
explicitly  known and the integration finally leads to
Eqs.(\ref{s20}-\ref{s23}).

\subsection{\label{app4} Calculation of
Onsager term of the Bose gas} First some elementary relations  are
deduced for later use. The definition $ n_\vk=\langle{\cal
N}_\vk\rangle_0$ leads directly to  $
\textrm{d}n_\vk/\textrm{d}\nu_\vk=\langle{\cal
N}^2_\vk\rangle_0-n_\vk^2$. From Eq.(\ref{b5}) we find
$\textrm{d}n_\vk/\textrm{d}\nu_\vk =n_\vk(1+n_\vk) $ and thus
\begin{equation}
\label{a100} \langle{\cal N}^2_\vk\rangle_0=n_\vk(1+2\, n_\vk)\:.
\end{equation}
 Generalizing this procedure to higher order we find
\begin{equation}\label{a101}
\langle{\cal N}^3_\vk\rangle_0=n_\vk(1+6\, n_\vk+6\, n_\vk^2)
\end{equation}
and
\begin{equation}\label{a102}
\langle{\cal N}^4_\vk\rangle_0=n_\vk(1+14\, n_\vk+36\,
n_\vk^2+24\, n_\vk^3)\:.
\end{equation}
Eq.(\ref{a100}) and the factorization property leads to
\begin{equation}\label{a105}
(\widehat{\cal N}_\vk|\widehat{\cal N}_{\vk'})_0= \delta_{\vk
\vk'}\:\langle\widehat{{\cal N}}_\vk^2\rangle_0= \delta_{\vk
\vk'}\:n_\vk(1+n_\vk)\:.
\end{equation}
 The
$\widetilde{{\cal N}}_\vk$ commute with ${\cal K}_0$. Therefore
$(\widetilde{{\cal N}}_\vk|{\cal U})_0=\langle\widetilde{{\cal
N}}_\vk{\cal U}\rangle_0$ holds. This implies that the projector
$\mathbb{P}_0$ defined by Eq. (\ref{17}) simplifies to
\begin{equation}\label{a106}
\mathbb{P}_0  {\cal U}= \langle{\cal U}\rangle_0 + \sum_{\vk}\,
\frac{\langle \widetilde{{\cal N}}_\vk {\cal
U}\rangle_0}{n_\vk(1+n_\vk)}\widetilde{{\cal N}}_\vk\:,
\end{equation}
where ${\cal U}$ is any   Hilbert space operator.

Let us introduce  a short hand notation by
 \begin{equation}\label{a120} {\cal B}_{\vp \vq
\vk}= \, {b}_{\vp+\vk}^\dag {b}_{\vq-\vk}^\dag {b}_{\vp}
{b}_{\vq}\;.
\end{equation}
Then the interaction Hamiltonian (\ref{b2int}) is rewritten as a
sum of two contributions
\begin{equation}\label{a121}
{\cal K}'= -\frac{\beta U}{2 V}{\cal Y}= -\frac{\beta U}{2
V}\left({\cal Y}^{(1)}+{\cal Y}^{(2)} \right)
\end{equation}
with
\begin{eqnarray}\label{a122}
{\cal Y}^{(1)} &=& 2\sum_{\vp  \neq\vq}{\cal N}_{\vp}{\cal
N}_{\vq}+\sum_{\vp}\left\{{\cal N}^2_{\vp}-{\cal N}_{\vp}\right\}
\quad\quad\quad\quad\\\label{a122b}
 {\cal Y}^{(2)}&=& \sum_{\vp\:\vq \:
 \vk\neq (0,\vq-\vp)}{\cal B}_{\vp \vq \vk}\quad.
\end{eqnarray}
With these definitions the Onsager term (\ref{20}) can  be
expressed as
\begin{equation}\label{a124}
G^{(2)}= \frac{\beta^2 U^2}{4 V^2}\left\{\left({\cal
Y}|\mathbb{Q}_0 {\cal Y}_{1}\right)_0+\left({\cal Y}|{\cal
Y}_{2}\right)_0\right\}\:,
\end{equation}
where we have already used that $\langle {\cal Y}_{2}\rangle_0=0$
and $\langle {\widetilde{{\cal N}}_\vk\cal Y}_{2}\rangle_0=0$ and
consequently $\mathbb{P}_0 {\cal Y}_{2}=0$ holds according to
Eq.(\ref{a106}).

The two terms of Eq.(\ref{a124}) will separately be treated. As
$\mathbb{Q}_0 {\cal N}_\vp{\cal N}_\vq=\widetilde{{\cal
N}}_\vp\widetilde{{\cal N}}_\vq$ results for $\vp\neq\vq $ we find
in consequence that the ${\cal N}_\vp $ commute with ${\cal K}_0$
\begin{eqnarray}
\nonumber  \left({\cal Y}|\mathbb{Q}_0 {\cal Y}_{1}\right)_0 &=&
2\sum_{\vp \neq\vq}\langle{\cal Y}\widetilde{{\cal
N}}_\vp\widetilde{{\cal N}}_\vq\rangle_0 +\sum_{\vp}\langle{\cal
Y} \mathbb{Q}_0{\cal N}^2_{\vp} \rangle_0
\quad\quad\quad\quad\\\nonumber
   &=&  8\sum_{\vp
\neq\vq}\langle\widetilde{{\cal N}}_\vp^2\widetilde{{\cal
N}}_\vq^2\rangle_0+\sum_{\vp}\langle{\cal N}^2_{\vp}
\mathbb{Q}_0{\cal N}^2_{\vp} \rangle_0\\ \nonumber&=& 8\sum_{\vp
\neq\vq}n_\vp(1+n_\vp)n_\vq(1+n_\vq)
\\ \label{a125}&{}&\quad\quad\quad\quad
+4\:\sum_{\vp} n_\vp ^2(1+n_\vp^2)
\end{eqnarray}
where in the last step Eqs.(\ref{a100} - \ref{a102}) are used.

With Eq.(\ref{a122b}) the second contribution to the Onsager term
is written as
\begin{equation}\label{a126corr}
\left({\cal Y}|{\cal Y}_{2}\right)_0 =\sum_{\vp\:\vq \:
 \vk\neq (0,\vq-\vp)}\left({\cal Y}
 |{\cal B}_{\vp \vq \vk}\right)_0\quad
\end{equation}
and we  focus on the calculation of $({\cal Y}
 |{\cal B}_{\vp \vq \vk})_0$ for $\vk\neq 0,\vq-\vp$ . Note that
\begin{equation}\label{a126}
 \left[{\cal K}_0\,,\,{\cal B}_{\vp \vq \vk}\right]=
 \sum_{\bar{\vk}} \nu_{\bar{\vk}} \left[{\cal N}_{\bar{\vk}}\,,\,
 {b}_{\vp+\vk}^\dag {b}_{\vq-\vk}^\dag {b}_{\vp} {b}_{\vq}\right]
 =\omega_{\vp \vq \vk}\,{\cal B}_{\vp \vq \vk}
\end{equation}
with
\begin{equation}\label{a127}
\omega_{\vp \vq
\vk}=\nu_{\vp+\vk}+\nu_{\vq-\vk}-\nu_{\vp}-\nu_{\vq}
\end{equation}
holds. Thus the relation (\ref{0a0}) can be employed that yields
\begin{eqnarray}
\nonumber
  \omega_{\vp \vq \vk}\:(\,{\cal Y}
 \,|\,{\cal B}_{\vp \vq \vk}\,)_0 &=&\langle \:[\,
 {\cal B}_{\vp \vq \vk}\,,\,{\cal Y}\,]\:\rangle_0
  \quad\quad\quad\quad\quad\quad\quad\quad\quad\\\nonumber
&=&
 4\,\langle \:[\,{\cal B}_{\vp \vq \vk}\,,\,
 {\cal B}_{\vp \vq \vk}^\dag\,]\:\rangle_0\\\nonumber
&=&4\:\big\{\:n_{\vp+\vk}\,n_{\vq-\vk}\:(1
+n_{\vp}+n_{\vq})\\\label{a128} &{}&-\:n_{\vp}n_{\vq}\:(1
+n_{\vp+\vk}+n_{\vq-\vk})\:\big\}
\end{eqnarray}
where the last step, the calculation of the commutator, is tedious
but straightforward.

With the definition (\ref{a127}) we find for $\vk\neq
0\,,\,\vq-\vp$
\begin{eqnarray}\nonumber
\left({\cal Y}|{\cal B}_{\vp \vq \vk}\,\right)_0 =4
\frac{n_{\vp+\vk}\,n_{\vq-\vk}\:(1 +n_{\vp}+n_{\vq})
}{\nu_{\vp+\vk}+\nu_{\vq-\vk}-\nu_{\vp}-\nu_{\vq}}\\\label{a129}
+4\:\frac{n_{\vp}n_{\vq}\:(1 +n_{\vp+\vk}+n_{\vq-\vk})}
{\nu_{\vp}+\nu_{\vq}-\nu_{\vp+\vk}-\nu_{\vq-\vk}}
\end{eqnarray}
where the $\nu_{\vp}$ as functions of the $n_{\vp}$ are given by
Eq.(\ref{b4}). Using this dependence we can calculate the limiting
behavior of  $\left({\cal Y}|{\cal B}_{\vp \vq \vk}\,\right)_0 $
for the excluded values of $\vk$ and we obtain
\begin{eqnarray}\label{a130corr}\nonumber
 \lim_{\vk\rightarrow0}\:\left({\cal Y}|{\cal B}_{\vp \vq
\vk}\,\right)_0 &=& 4n_\vp(1+n_\vp)n_\vq(1+n_\vq)\\\nonumber
\lim_{\vk\rightarrow\vq-\vp }\:\left({\cal Y}|{\cal B}_{\vp \vq
\vk}\,\right)_0&=&4n_\vp(1+n_\vp)n_\vq(1+n_\vq)\;
\end{eqnarray}
for both cases. These findings imply that   just  one unrestricted
triple sum  remains
\begin{eqnarray}
 \label{a130}
 G^{(2)}&= &\frac{2\beta^2 U^2}{ V^2}\sum_{\vp\:\vq \:
 \vk}
\frac{n_{\vp+\vk}\,n_{\vq-\vk}\:(1 +n_{\vp}+n_{\vq})}
{\nu_{\vp+\vk} +\nu_{\vq-\vk}-\nu_{\vp}-\nu_{\vq}} \quad\quad
\end{eqnarray}
 and all the other contributions cancel out.



\begin{thebibliography}{20}
\bibitem{com1} For some applications it may be convenient to work with complex
parameters and non-Hermitian  operators. The generalization to
this case is straightforward.
\bibitem{comast}  The usual product ${\cal U}{\cal V}$
 of two operators is defined
via the action on  vectors of the unitary space. No such action is
defined for  ${\cal U}\ast{\cal V}$.
\bibitem{comhubb}
  The expansions performed for the Hubbard model
\cite{gyhub} do not satisfy this requirement. Thus our approach
does not include this work.
\bibitem{ww} The Dzyaloshinskii-Moriya interaction is not included.
 A generalization to this case is straightforward.
\bibitem{diff} The argumentation for the quantum case is  not as strict as
for the classical case. According to Eqs.(\ref{400}) and
(\ref{s1}) the exact relation
\begin{equation}\nonumber
 \partial_\alpha G_\alpha=
\frac{\beta}{2 } \sum_{i,j}J_{ij}\;
\bm{m}_{\,i}\cdot\,\bm{m}_{\,j} +\frac{\beta}{2
}\sum_{i,j}J_{ij}\;
\langle\bm{\widetilde{s}}_{\,i}\cdot\,\bm{\widetilde{s}}_{\,j}\rangle_\alpha
\end{equation}
holds. For classical systems the susceptibility is given by
$\chi_{ij}^{\mu\hat{\mu}}=
\beta\langle{\widetilde{s}}_{\,i}^{\,\mu}\,
\,{\widetilde{s}}_{\,j}^{\,\hat{\mu}}\rangle_\alpha $ , and
singularities of $\chi_{ij}^{\mu\hat{\mu}}$ imply singularities of
$\partial_\alpha G_\alpha $ and consequently of $ G_\alpha $.

For quantum systems this direct conclusion does not apply as
$\chi_{ij}^{\mu\hat{\mu}}=\beta({\widetilde{s}}_{\,i}^{\,\mu}\,|
\,{\widetilde{s}}_{\,j}^{\,\hat{\mu}})_\alpha $ holds which
differs from $\beta\langle{\widetilde{s}}_{\,i}^{\,\mu}\,
\,{\widetilde{s}}_{\,j}^{\,\hat{\mu}}\rangle_\alpha $.
\bibitem{second}
To be more detailed we consider Eq.(36) of \cite{baym} with
$z_\nu+\mu'= \epsilon^0_\vk $ and replace already in this equation
of \cite{baym} all the bare energies $\epsilon_\vp^0 $ by the
dressed energies $\hat{\epsilon}_\vp $. The resulting equation
agrees with Eq.(\ref{b10}) as the variables $\nu_\vp $  are
related to the $\hat{\epsilon}_\vp $ via $\nu_\vp=\b(\mu'-
\hat{\epsilon}_\vp )$, according to Eq.(\ref{b18}).
\bibitem{care}
 From the experience for spin glass systems \cite{I,asp} (and
compare Sec.\ref {secpara}) such arguments should be used with
some care for projector type contributions like the first order
terms $U/V$. For positive values of $U$, however, the protector
term has no influence to the lower end of the spectrum of
$\chi_{\vk \vk'}$.
\end{thebibliography}
\end{document}